\documentclass[fleqn,usenatbib]{mnras}

\usepackage{newtxtext,newtxmath}

\usepackage[T1]{fontenc}

\DeclareRobustCommand{\VAN}[3]{#2}
\let\VANthebibliography\thebibliography
\def\thebibliography{\DeclareRobustCommand{\VAN}[3]{##3}\VANthebibliography}

\usepackage{graphicx}	
\usepackage{amsmath}	
\usepackage{gensymb}
\usepackage{siunitx}
\usepackage{booktabs}
\usepackage{pdflscape}
\usepackage[table]{xcolor}
\usepackage{xcolor}
\newcommand{\update}[1]{#1}

\usepackage[normalem]{ulem}

\title[Astrometric Jitter Measurements]{Astrometric exoplanet detection survives solar-like stellar
contamination}

\author[C. Deagan et al.]{
Conaire Deagan,$^{1}$\thanks{E-mail: C.Deagan@unsw.edu.au}
Benjamin T. Montet,$^{1}$
P. Tuthill,$^{2}$
M. Ferraro,$^{1}$
R. Lyu,$^{1}$
and E. Sheehan.$^{1}$
\\
$^{1}$School of Physics, University of New South Wales, Sydney, NSW 2052, Australia\\
$^{2}$University of Sydney, Sydney, NSW, Australia
}

\date{Accepted XXX. Received YYY; in original form ZZZ}

\pubyear{\the\year{}}

\begin{document}
\label{firstpage}
\pagerange{\pageref{firstpage}--\pageref{lastpage}}
\maketitle

\begin{abstract}
Astrometric monitoring of stars provides a promising method for discovery of low-mass planets around nearby Sun-like stars. The astronomical community has proposed several telescopes designed to perform high-precision astrometric observations. One limiting factor intrinsic to stars is the astrometric noise - or "jitter" - induced by surface stellar activity such as starspots and faculae. Despite previous estimates, the relative size of this signal  has not been empirically measured from direct photometric observations. 
We analyse high-resolution images of the Sun to quantify the photometric centroid jitter across three narrow wavelength regions over nearly a decade, spanning high and low activity periods of the Solar cycle. We compare our findings to previous theoretical estimates.
We scale this jitter to simulate how a Solar-twin would appear at various distances, establishing an astrometric noise floor below which detection is significantly complicated by stellar activity. We also introduce starspot simulations that augment our data. We find the typical astrometric jitter of the Sun at  \(\lambda = 607.2 \pm 0.25\text{nm}\) to be \(0.342\mu \text{as pc}\), ranging between \(0.058\mu \text{as pc}\) and \(1.294\mu \text{as pc}\) for low and high activity periods, respectively. This is lower than the expected \(\approx 3\mu \text{as}\) astrometric signal that an Earth-like planet would produce around a Sun-like star, at 1 pc. Therefore, the astrometric noise floor imposed by intrinsic stellar activity sets a detection limit below one Earth but greater than Mars around Solar-analog stars, making instrument precision the limiting factor for Earth-like exoplanet searches.
\end{abstract}




\section{Introduction}
\begin{figure}
\centering
\includegraphics[width=0.95\linewidth]{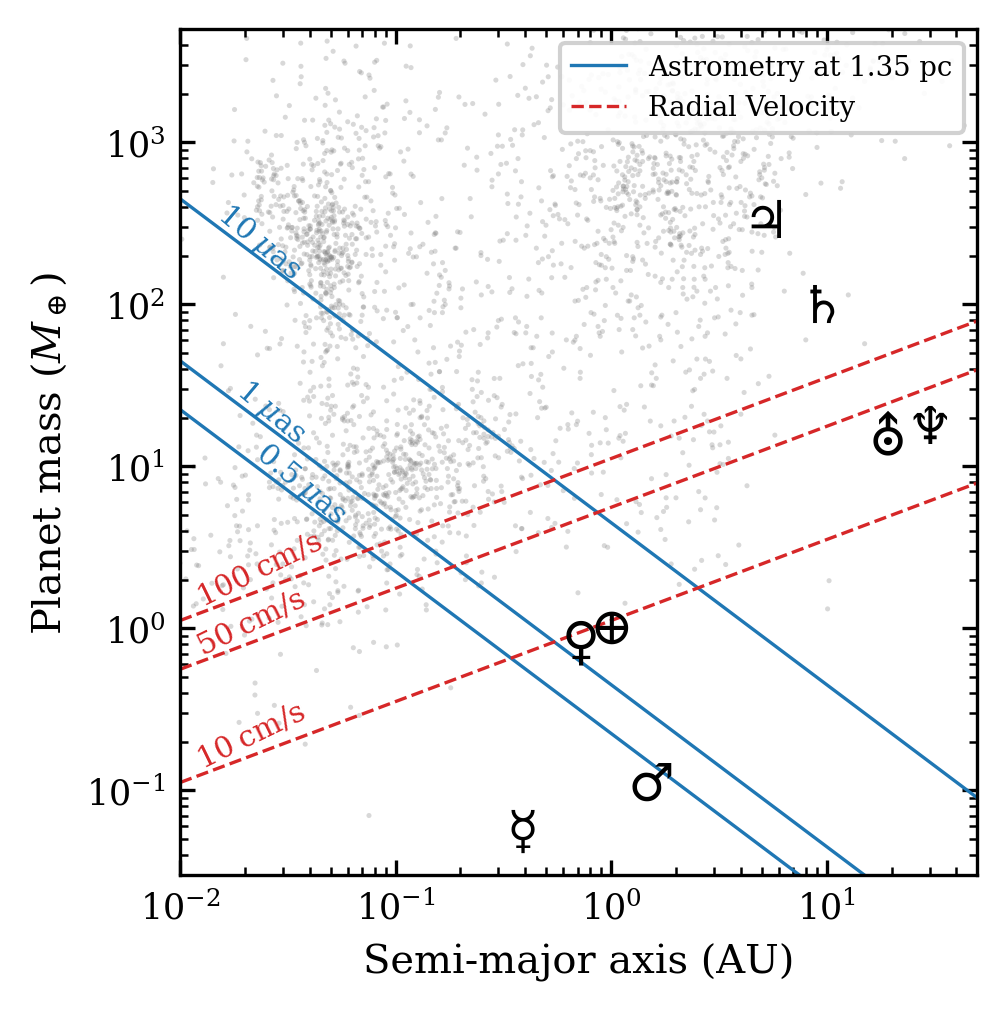}
\caption{A comparison between exoplanet detection using RV and astrometry for a planet on a circular, edge-on orbit around a 1 $M_\odot$ star. The position of the Solar System planets have also been marked. This work calculates the astrometric signal at a distance chosen to align with Alpha Centauri \update{The grey dots indicate confirmed exoplanet parameters only; their positions do not reflect detection signal strength.} Both RV and astrometric signals increase with planet mass, but unlike the RV signal, the astrometric signal increases with orbital separation.}
\label{fig_LinesOnMap}
\end{figure}

Precise astrometric measurements of stellar positions in the sky allows the detection of the gravitational reflex motion induced by an orbiting exoplanet, as the star orbits the common barycentre \citep{Perryman2014}). As of August 2025, only 5 of the 5983 confirmed exoplanets have been detected via astrometry and all five are several times the mass of Jupiter (\cite{Sahlmann2013}, \cite{Curiel2022}, \cite{Sozzetti2023}, \cite{Stefnsson2025}). Although the Gaia mission is expected to find tens of thousands of exoplanets via astrometric monitoring of their host stars, most of these planets will be more massive than Jupiter (\cite{Perryman2014}). While current astrometric telescopes like Gaia lack the (sub-)microarcsecond precision required to detect rocky planets, many telescopes that aim to meet this requirement have been proposed, such as Theia (\cite{TheiaPaper}), small-Jasmine (\cite{Utsunomiya2014}), MASS (\cite{Nemati2020}), SIM PlanetQuest (\cite{Catanzarite2006}), NEAT (\cite{NEAT}), CHES (\cite{Ji2022}), AMARDA \citep{Gardner2022}, and TOLIMAN (\cite{Tuthill2018}). Of these, only a few have been funded and none are currently operational. Additionally, two different high-precision astrometric technologies are currently under development (\cite{Bendek2021}). Despite the lack of current detections, there are excellent reasons to explore the use of high-precision astrometry to detect exoplanets. 
Namely, the lower intrinsic noise of astrometric measurements for nearby stars (quantified in this work) enables superior signal-to-noise ratios in parameter space regions where other traditional methods, such as Doppler RV, become noise-dominated (as discussed below). Microlensing presents an exception, as it is also sensitive to low-mass planets on wide orbits. However, it has the fundamental limitation of relying on chance, one time events - meaning that follow up observations are generally not possible. (\cite{Gaudi2012}).
Unlike most other exoplanet detection methods, the strength of the astrometric signal {\it increases} with orbital separation, as demonstrated in Equation (\ref{eq_astrometry_signal}) (\cite{Perryman2018}).

\begin{equation}
    A = \left(\frac{M_p}{M_*}\right) \left(\frac{a}{1\text{AU}}\right) \left(\frac{1 \text{pc}}{d}\right) \text{arcsec}  
\label{eq_astrometry_signal}
\end{equation}

where \(A\) is the astrometric signal in arcseconds, \(Mp\) and \(M_*\) are the masses of the planet and star respectively, \(a\) is the semi-major axis of the planets orbit, and \(d\) is the distance to the star from Earth. In contrast, the gravitational reflex motion of a star induced by a planet on a circular orbit will generate a radial velocity (RV) signal with a semi-amplitude as described in Equation (\ref{eq_rv_signal}) (\cite{Perryman2018}).

\begin{equation}
 K = \frac{1}{\sqrt{a}} \frac{\sqrt{G}}{(M_* + M_p)^{2/3}} \frac{M_p \text{sin}i}{\sqrt{1-e^2}}
 \label{eq_rv_signal}
\end{equation}
where \(K\) is the RV signal, \(G\) is the gravitational constant, \(e\) is the eccentricity of the planets orbit and \(i\) is the inclination of the planets orbital plane relative to us. This difference is shown in Fig. \ref{fig_LinesOnMap}, which demonstrates the astrometry and RV signals as a function of orbital separation and planet mass. For an Earth analogue at \(\approx 1\text{pc}\), the RV signal is \(\sim 10\mathrm{cm s^{-1}}\), while the astrometric signal is on order of a few micro-arcseconds, depending on the distance of the system from Earth.
Very few RV instruments can achieve the precision required to detect an Earth twin. 
VLT/ESPRESSO, a state of the art instrument with an instrument precision of \(10 \mathrm{cm s^{-1}}\), can only just begin to detect the signal of an Earth twin in the absence of stellar noise, although on-sky performance has only shown a precision of better than \(25\, \mathrm{cm s^{-1}} \) \citep{Pepe2021}. While RV methods have achieved impressive precision and are now primarily limited by stellar noise modeling, astrometry currently operates in a different regime where instrumental improvements can still yield significant gains before stellar noise becomes the dominant limitation.

The variations induced by stellar activity impose fundamental physical limits in signal recovery for both techniques. This signal comes from three main sources: stellar oscillations (mainly p-mode waves), granulation, and magnetic activity, such as starspots \citep{Howard2012}. Stellar oscillations can generate noise on the order of \(1 \mathrm{m s^{-1}}\) for solar-like stars,  and granulation can generate noise levels of up to several 10's of \(\mathrm{cms^{-1}}\) when disk integrated  \citep{Yu2018,Cegla2019}. Both phenomena occur on relatively short timescales - ranging between a few minutes to a few tens of minutes  (\cite{Borgniet2015}). \update{On longer timescales, supergranulation can generate RV noise on the order of several 10's of \(\mathrm{cms^{-1}}\), which can be limit exoplanet detection rates significantly \citep{Meunier2019SG}.} Stellar magnetic activity, such as starspots and faculae, can also generate significant RV shifts, on a rotational or activity cycle timescale \citep{Borgniet2015}. The exact level that magnetic features affect its RV signal depends heavily on a star's activity cycle \citep{Borgniet2015}, and the star itself \citep{Saar1997}. Many techniques exist that aim to reduce this RV jitter. For example, \cite{Dumusque2018} analyse the RVs of individual lines in high-resolution spectra, leveraging the fact that certain spectral lines are more sensitive to stellar activity than others to partially separate stellar activity from planetary signals, demonstrating a 1.6× reduction of red noise in the 2010 alpha Cen B RV measurements. Gaussian Processes (GPs) are a frequently adopted method, with \cite{Rajpaul2015} demonstrating their ability to recover planetary signals even in challenging cases where the planetary signal is weaker than the activity signal and both share identical periods.  Although several techniques to minimise and remove this stellar signal exist, current methods have not yet reliably reduced the level of RV jitter down to a level where they could detect an Earth-twin (which would generate a signal of \(\sim 10 \mathrm{cms^{-1}} \)), with stellar activity limiting RV precision to  \(\gtrsim 1\mathrm{ms^{-1}}\) \citep{deBeurs2022} \citep{Cretignier2022}.

Given the limitations with traditional RV exoplanet detection, a few have proposed high-precision astrometry as an alternative \citep{TheiaPaper, Ji2022}. However, one astrophysical obstacle is that the astrometric jitter of the Sun has not been reliably measured using Sun-as-a-star photometric methods, so estimations of the jitter of Sun-like stars may be inaccurate. If the astrometric jitter was notably larger than the planetary astrometric signal, then astrometric approaches would face similar challenges to RV methods. Several papers have published estimates and/or measurements of astrometric jitter (see \cite{Catanzarite2008}; \cite{Makarov2010}; \cite{Lagrange2011}; \update{\cite{Meunier2020}}; \cite{Shapiro2021}; or \cite{KaplanLipkin2022})), but these are often simulation based, or are measured using non-photometric data.  Additionally, there is a mild tension present as the different measurements of the average jitter do not all mutually agree. To quantify this discrepancy, we note that \update{\cite{Meunier2020}} reports typical astrometric jitter of \(1.06\mu\text{as pc}\) while \cite{Makarov2010} reports \(0.65\mu\text{as pc}\) for similar solar conditions.


In this paper we present an analysis of data from the Precise Solar Photometric Telescope (PSPT) at Mauna Loa Solar Observatory (MLSO) and quantify the astrometric jitter of the Sun-as-a-star in multiple wavelengths, and we discuss our results in comparison to previous works. This paper is structured as follows: In Section \ref{sec:data} we describe the PSPT dataset and how we transform it. In Section \ref{sec:methods}, we describe how we made our independent measurements of the astrometric jitter. Section \ref{sec:results} discusses the results of our analysis. In Section \ref{sec:discussion} we discuss the progression of previous estimates of the astrometric jitter, and why they differ from ours. Additionally we discuss for what purposes both our results and results in the literature should be used. We conclude in Section \ref{sec:conclusion}.

\section{ Data}
\label{sec:data}
We use data from the Precision Solar Photometric Telescope \cite[][PSPT]{White2000}, which is a part of the Radiative Inputs from the Sun to the Earth (RISE) project (\cite{RISE}). There are three such telescopes: one in Rome at the Osservatorio Astronomico di Roma, one in Hawaii at MLSO, and one in New Mexico at the National Solar Observatory (see \cite{Ermolli2022} for details). Our data comes solely from MLSO/PSPT primarily due to the accessibility of the data, but also as this location has the highest resolution of 1 arcsecond per pixel, with images of the full solar disk containing 2k x 2k pixels (\cite{Vogler2005}). The per-pixel photometric precision is \(0.1\%\). Via a bootstrapping analysis, we find that this photometric uncertainty leads to a per image photocentroid uncertainty of approximately 100 pico-arcseconds. The dataset used in this paper covers three narrowband filters, `blue' (\(\lambda = 409.3\pm0.15nm\)), `red' (\(\lambda = 607.2 \pm 0.25nm\)), and the activity indicator line of `Ca~{\sc ii} K' (\(\lambda = 393.4\pm0.15nm\)), and spans a period of 10 years (2005-2015) covering the end of Solar cycle 23 and most of solar cycle 24. Although the instrument has measured other wavelengths and has been collecting data since 1998, we chose to use the aforementioned subset to ensure a consistent, well-sampled, and uniform dataset, as data outside this range is less reliably uniform\footnote{see \url{https://lasp.colorado.edu/pspt_access/} for details.}. This results in a dataset of approximately 12,500 images each for the red and blue filters, and approximately 25,500 images for the Ca~{\sc ii} filter.

\begin{figure}
\centering
\includegraphics[width=0.95\linewidth]{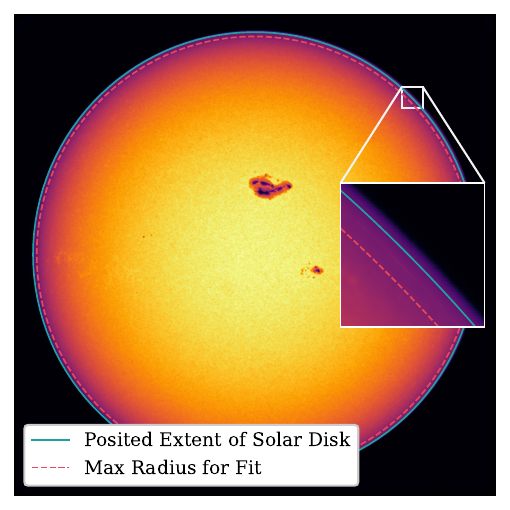}
\caption{A demonstration of the limitations of the PSPT data processing pipeline, which did not process the limb uniformly with the rest of the Solar disc. The solid blue line is the stated extent of the solar disk in the original dataset, and the dashed red line is the extent to which the data has been cleaned. Note between the red and blue lines there exists a slight intensity increase --- an example of the difference between the processed and unprocessed data separated by this boundary.}
\label{fig_PictureInPicture}
\end{figure}

\section{ Methods}
\label{sec:methods}
\subsection{Data Quality}
The available images were preprocessed to remove instrumental and non-physical signatures, including dark current, gradients, and quadrant artifacts. Corrections for these effects were applied outward from the geometric centre of the solar disk to nearly the full solar radius. The median ratio of the corrected radius to the full radius was 0.934, with 95\% of values between 0.924 and 0.931. Beyond this region, instrumental artifacts remain (see Fig. \ref{fig_PictureInPicture}), rendering the outer areas unsuitable for photocentroid calculations.
Given the large number of images, it was infeasible to check the quality of each one. We performed automated checks to ensure the integrity of the data. These included:
\begin{enumerate}
    \item Randomly sampling 200 images per wavelength to establish typical bounds for statistical measures of the solar intensity distribution.\footnote{That is, the distribution of pixel intensities of the solar disk.} These measures included the median, mean, standard deviation, maximum intensity, kurtosis, and integrated intensity (sum). For each metric, the acceptable range spanned \(\pm3\) standard deviations from the median value. Any image with a value outside these bounds was flagged. 
    \item Ensuring the solar disk was at least 50 pixels from all detector edges. 
    \item Rejecting images containing NaN values or significantly negative pixel values \((<-1 \text{ counts})\).
\end{enumerate}
We noted the time of observation for any image that was flagged as potentially invalid, and then all images taken that day across the three wavelengths were manually checked to ensure that no corrupted data was included. 
Finally, after all appropriate transformations to the data were made (described below), any recovered astrometric measurement more than 3 standard deviations from the mean was manually checked to ensure that any other problematic images were removed. Only a few hundred images were outside this range. The manually checked images did sometimes contain physical, non-Solar contaminants, such as the 2012 transit of Venus and a partial solar eclipse. These images were removed.

\subsection{Full-disk scaling factor}
\label{sec:fulldisk}
Our analysis requires accounting for the fact that PSPT images are cropped at the solar limb, excluding a narrow annulus of data that could influence photocentroid calculations through lever-arm effects. While this cropped region is small, we developed a scaling factor to correct for this systematic bias.

\begin{figure}
\centering
\includegraphics[width=1.05\linewidth]{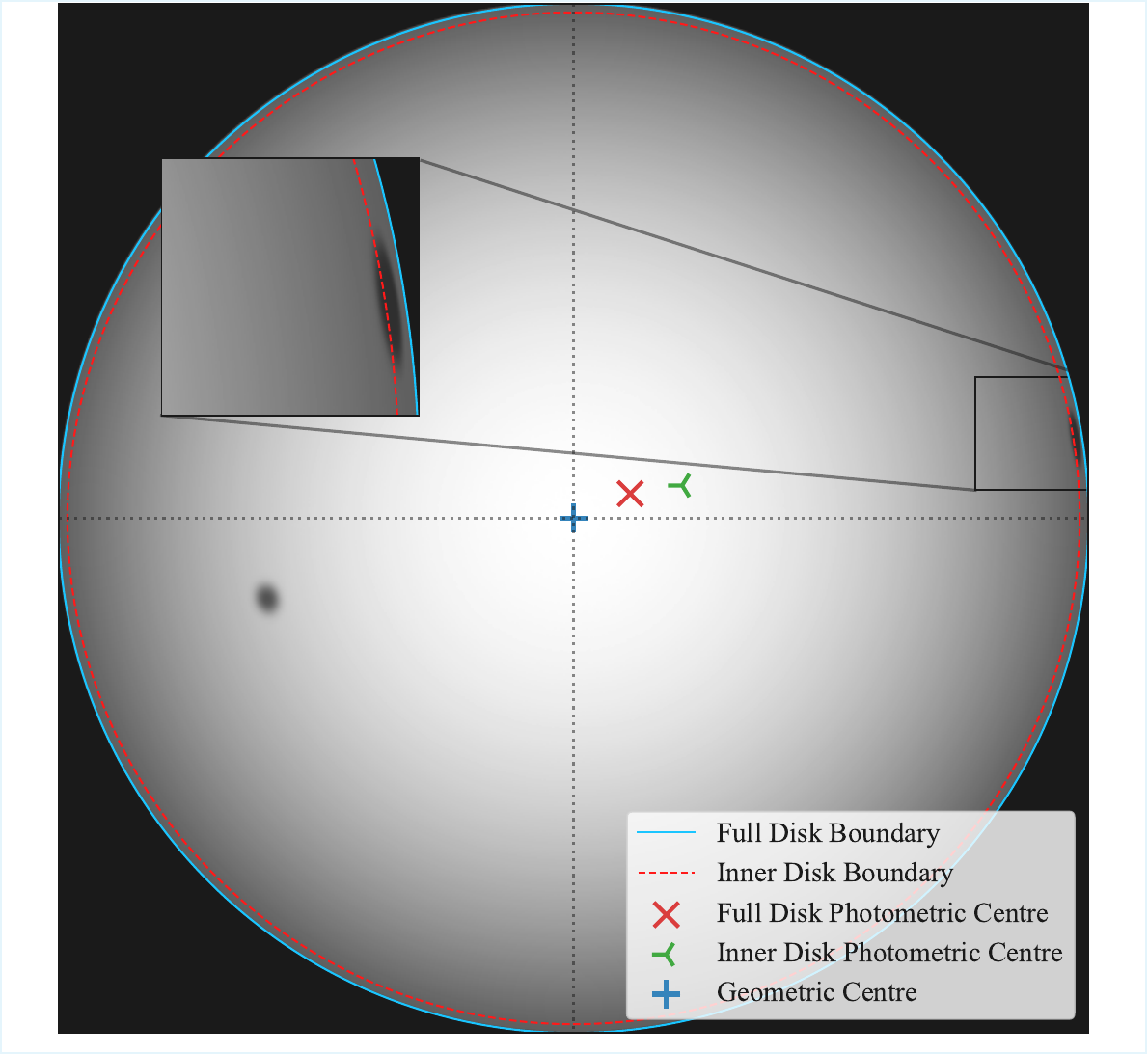}
\caption{The impact that a cropped measurement has on the photometric centre of the solar disk. Here, the inner disk boundary is at 98.3\% of the full radius - the median value of the cleaned/full radii ratio in the PSPT dataset. The image shows two simulated Sunspots, one in the lower left quadrant, and one in the upper right quadrant on the limb of the disk. This spot is heavily foreshortened. A typical Solar quadratic limb darkening has been applied. The photometric centre of the full disk is marked by the red cross, and the photometric centre of the cropped disk is marked by the green triaxial symbol. The positions of these points relative to the geometric centre have been exaggerated by a factor of 1000. It is clear that as most of the upper right Sunspot has been excluded from the inner disk, its impact to deflect the signal has been suppressed (and thus the relative influence of the lower left spot is greater, moving the green triaxial marker further away from the geometric centre). In this contrived example, the photocentre displacement is roughly doubled. However, the jitter depends on the distribution of many such displacements rather than any individual case (as further discussed in the text).}
\label{fig_SimPictureInPicture}
\end{figure}

To determine this scaling factor, we created a sunspot model to determine the ratio between cropped and full-disk photometric deflections. While a single scaling factor cannot capture the full effect of the cropped disk---as the cropped region can either increase or suppress photometric deflection depending on spot configuration (see Figure \ref{fig_SimPictureInPicture})---a complete treatment would require recreating the entire PSPT dataset with detailed, time-varying, evolution of the solar surface. This is feasible in principle but unnecessary, since the dominant contribution to jitter arises from the stochastic nature of individual spot appearances and locations rather than their detailed temporal evolution.

The model used circular spots with Gaussian blur and solar-like spot contrast ($\sim$0.7), with sizes sampled from a realistic log-normal distribution \citep[\(\langle A\rangle = 58.6,\,\, \sigma_A=2.49\) micro solar hemispheres;][]{Baumann2005}. Spots were placed at latitudes sampled from a double Gaussian straddling the equator (standard deviation 5°, means varying $\pm$10--20° to emulate solar cycle latitude migration), with spot numbers drawn from the smoothed monthly sunspot record \citep{SILSO_Sunspot_Number}. Over a simulated 10-year period at multiple inclinations, we calculated photometric centres for both full and cropped disks. The resulting scaling factor was 0.98 \(\pm\) 0.01, indicating that the outer limb typically suppresses photometric deflection by $\sim$2\%.  Less than 1\% of the time is this scaling factor greater than one. This marginal effect is dwarfed by the quasi-stochastic appearance of surface magnetic features.

\subsection{Measuring Astrometric Jitter}
As the Earth orbits the Sun, both the portion of the Sun we observe and its apparent size change. These variations are captured in three parameters from the PSPT dataset: SOLAR\_P0, SOLAR\_B0, and SOLAR\_R. SOLAR\_P0 represents the angle between the geocentric north pole and the solar rotational north pole, varying between \(\pm 26.3\) degrees throughout the year. SOLAR\_B0 is the heliographic latitude of central point of the solar disk, resulting from the tilt between the ecliptic and the solar equatorial plane, ranging between \(\pm 7.23\) degrees. SOLAR\_R reflects the apparent solar radius, which varies by approximately 30 pixels due to Earth's elliptical orbit.

Astrometric jitter is the variation in the photocentroid. To calculate this, we created a circular mask centred on the geometric centre of the Sun, based on coordinates provided in the image header, which extended to the radius that each image was cleaned. For each masked image, we calculated the flux-weighted centre of mass -- this is the `photocentroid'. We calculated the x and y components (in the image reference frame) of the photocentroid deflection by subtracting the position of the photocentroid from the geometric centre.  This residual is the jitter vector. We rotated this vector by the negative of the SOLAR\_P0 angle to align the x and y components with the equatorial and polar axis of the Sun. We converted this vector to a physical distance by dividing by the apparent radius of the Sun in pixels. We applied a scaling factor to account for the overall effect of the annulus of unprocessed data. Although the width of this annulus is small -- on the order of 10's of pixels -- this region has a lever-arm effect on the centre of mass calculation. When a spot appears in the processed region of the disk, this annulus suppresses the photocentroid deflection. Although the magnitude of this suppression depends on the number and location of surface magnetic features, we used only a single scaling factor as we are only interested in the statistical properties of the astrometric jitter. 

We did not correct for the SOLAR\_B0 angle variation, which changes the heliographic latitudes visible from Earth throughout the year. At certain times of the year, the region around one of the poles will not be visible. To correct for this would require knowledge of the surface of the Sun outside the visible hemisphere at the time.  However, this limitation has minimal impact on the photocentroid deflection as these polar regions lie well outside the active latitudes where large-scale magnetic features typically occur. To confirm this, we analysed the recorded latitudes of sunspots present in the Royal Greenwich Observatory (RGO) and United States Air Force (USAF) / National Oceanic and Atmospheric Administration (NOAA) sunspot records (\cite{rgo_msfc_sunspot}), finding that no sunspots were recorded in the top and bottom \(7.5\) degree latitudinal bands in the years between 1874 and 2013. \update{While facular and network elements do extend to higher latitudes, the observer always views at least one polar region regardless of B0, so the variation shifts the balance between the two polar contributions. As shown by \cite{Shapiro2021} and \cite{Makarov2010}, the annual variation in heliographic latitude manifests as a yearly modulation of the baseline north-south photocentre position rather than a change in jitter amplitude. This effect is unique to observations of the Sun and would not be present for other stars. Following \cite{Makarov2010}, we performed a periodogram analysis on our $\Delta y$ time series for the blue waveband and determined that the best fitting sinusoid at a 1-year period has an amplitude of $0.086\mu \text{as pc}$. We also simulated 250 months of solar-like spot configurations (using the spot model described in Section \ref{sec:fulldisk}) for observer heliographic latitudes between $-8\degree$ to $+8\degree$ in $1\degree$ increments and computed the median polar jitter at each latitude. Taking the median polar jitter at each latitude and the $1\sigma$ spread of photocentre, we used nested sampling \citep{Skilling2004} and calculated the Bayesian evidence for a constant, linear, and quadratic dependence on latitude. Under equal model priors, the constant is the weakly preferred model, with a posterior probability of $\sim68\%$, compared to the $\sim18\%$ for the quadratic and $\sim14\%$ for the linear model. We conclude that the the observed $B_0$ variation does not come from spots, but rather it possibly arises from large scale facular and network elements.}

\subsection{Correlations between Photocentroid Offset and Total Solar Irradiance}
\label{sec:correlations}
\update{We quantify the relationship between photometry and photocentroid motion by comparing the astrometric data derived from the PSPT dataset with contemporaneous Total Solar Irradiance (TSI) measurements from VIRGO/SOHO \citep{Finsterle2021}. The VIRGO TSI record span 1996-2021, fully covering the PSPT baseline. Only overlapping dates are used. To align the measurements between the datasets, the photometric offset and TSI are linearly interpolated and resampled at a daily cadence. To measure correlations in variability amplitude, we compute 30-day rolling standard deviations for TSI and photocentre deflections. The impact of window size is shown in Figure \ref{fig:windowsize}.
 This standard deviation-based metric summarises activity on rotational time scales and is less sensitive to short-term sign reversals and other sources of statistical noise. } 


\subsection{Non-photosphere sources of photocentroid deflection}
\label{sec:complications}
Thus far, we have discussed the impacts of surface stellar features --- extending to nearly the edge of the stellar disk --- on photocentroid calculations. However, there are additional potential causes of photometric centroid deflection to consider. We will discuss four such things here: prominences, comets, reflected starlight from planets, and zodiacal dust. 

\subsubsection{Prominences \& the corona}
The most notable solar phenomenon excluded from our analysis due to PSPT's limited clean region are prominences. These structures, which are cooler than the corona and less dense than the photosphere, exhibit negligible blackbody radiation and are primarily observable through active spectral lines \citep{Zhou2021}. While direct quantification of prominence contributions to photocentroid jitter is beyond the scope of our dataset, their continuum flux is lower than even the limb-darkened photosphere. Although prominence effects would be amplified by the lever-arm effect of the photocentroid calculation, typical prominence heights do not expand significant distances beyond the limb. However, prominence contributions would likely be enhanced in active wavelengths due to increased emission in spectral lines, providing additional motivation for avoiding magnetically sensitive passbands in astrometric mission design. Given these dataset limitations, we cannot definitively constrain prominence effects and recommend this as a direction for future investigations using extended solar imaging datasets.

The corona, likewise excluded from PSPT's cleaned region, cannot be quantified for photocentroid impact with this dataset. Constraining coronal effects would likely require coronagraph observations, such as from SOHO/LASCO \citep{Brueckner1995}, and require alignment to photospheric imaging. While coronal brightness is $<10^{-6}$ that of the photosphere \citep{Hanaoka2012}, quantifying its contribution to solar astrometric jitter is valuable for future work.

\subsubsection{Reflected Starlight}
Although planets reflect only a small fraction of starlight, their contribution could, in principle, shift the photocentre due to the lever-arm effect. For an unresolved star–planet system, the observed flux is
\begin{equation}
    F_{obs} = F_* + F_p    
\end{equation}
Where $F_{obs}$ is the total observed flux, $F_*$ is the stellar flux, and $F_p$  is the flux reflected by the planet. Neglecting planetary thermal emission (assumed to peak in the infrared), the reflected flux can be expressed as:
\begin{equation}
    F_p = \frac{1}{4\pi a^2} F_* A \phi(t)
\end{equation}
here $a$ is the distance from the star to the sun, $A$ is the Bond albedo, and $\phi(t)$ is the phase function describing the illuminated fraction of the planetary disk as seen by the observer. In the limiting case of perfect reflection \((A = 1)\) and full illumination of a planet with radius $r$, this reduces to
\begin{equation}
    \frac{F_p}{F_*} = (\frac{r}{2a})^2
\end{equation}
The resulting photocentre displacement scales as
\begin{equation}
    A_p \propto \frac{r^2}{4a}
    \label{scalesas}
\end{equation}
where \(A_p\) is the photometrically induced photocentre shift. 
\begin{figure}
\centering
\includegraphics[width=1.0\linewidth]{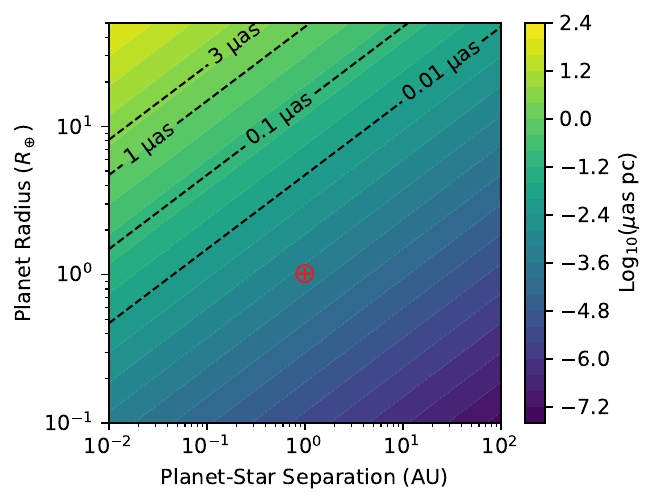}
\caption{The photocentroid deflection induced by reflected starlight from planets. The effect induced by Earth - assuming a maximal albedo and phase - has been marked.}
\label{fig:reflectedstarlight}
\end{figure}
Figure \ref{fig:reflectedstarlight} demonstrates that reflected light produces a negligible photocentre shift for most planetary radii and separations. Only hot Jupiters approach detectability, and even then, adopting Jupiter’s measured Bond albedo of 0.34 \citep{Hanel1981} reduces the deflection by a factor of \(\sim 3\) smaller. The signal is further diminished away from full phase and varies coherently with the orbital period, making it potentially distinguishable. Further work is needed to investigate the full impact of this effect on parameter estimations of the orbiting body. There may, for example, be a degeneracy between albedo and planetary mass. 
\subsubsection{Comets}

Sungrazing comets are a class of bright comets that pass close to the Sun, typically within 3.5 solar radii of the solar surface \citep{Jones2017}.  \update{We calculated the photocentroid deflection comets would induce using Equation \ref{scalesas}, which relates the observed jitter to the flux ratio between the comet and star and their separation. For our calculations, we treated the comets as point sources.} Our calculations convert the observed magnitudes to flux ratios relative to the Sun, then compute the resulting angular deflection at 1 pc distance.

The vast majority of sungrazers detected by space-based coronagraphs like SOHO are small objects with nuclear radii of up to 50 m \citep{Knight2010}, though most of their observed brightness comes from the surrounding dust coma rather than the nucleus itself \citep{Jones2017}. The data shown in Figure 7 comes from Table 2 of \cite{Knight2010}, which lists peak magnitudes for 65 Kreutz comets observed by SOHO's C3 coronagraph using a clear filter with bandpass 4000-8500 \r{A}. These comets typically reach peak magnitudes of around 5 and brighten dramatically as they approach the Sun, following power laws of $r^{-3.8\pm0.7}$ from 16-24 solar radii and $r^{7.3\pm2.0}$ beyond 24 solar radii \citep{Knight2010}.

\begin{figure}
\centering
\includegraphics[width=1.05\linewidth]{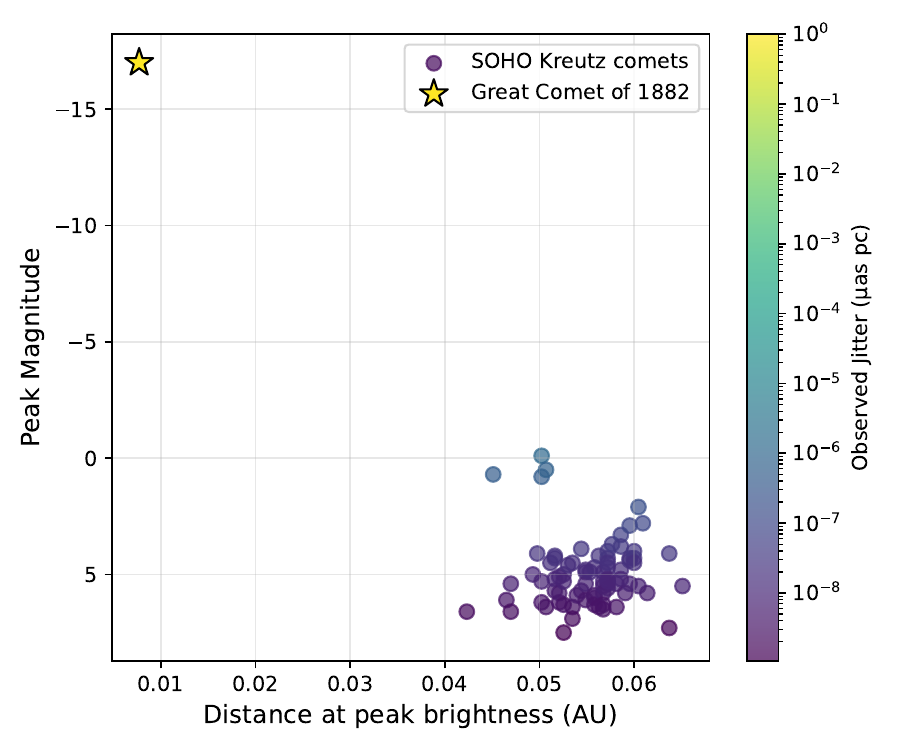}
\caption{The photocentroid deflection induced by reflected starlight from comets. The contours show the logarithm of the photocentroid deflection in $\mu$as pc. Data for comet brightnesses and perihelion distances are from \citet{Knight2010}, who analysed photometry of over 900 Kreutz sungrazers observed by SOHO from 1996--2005. The Great Comet of 1882 has been marked as an extreme historical example. }
\label{fig:sungrzer}
\end{figure}

For our calculations, we treated the comets as point sources, though in reality \cite{Knight2010} used circular apertures of 4 pixels radius for C3 images, corresponding to 43.8 arcmin$^2$, with the C3 coronagraph having a pixel scale of 56 arcsec/pixel. Even if we conservatively assume the entire aperture has the brightness reported in their table, the resulting photocentroid deflection for typical SOHO-observed sungrazers would still be many orders of magnitude below detection thresholds of \update{proposed high-precision astrometric missions such as TOLIMAN} (see Figure \ref{fig:sungrzer}).

Exceptional historical cases exist --- for example the Great Comet of 1882 reportedly reached a magnitude of between -10 and -17 for several hours near perihelion at 0.0078 AU \citep{ridpath2012great, Sekanina2007}. For such an event, our calculations indicate the photocentroid deflection could approach 1 microarcsecond, though this would persist for only hours. These great comets are exceedingly rare, with only a handful of them observed from the ground since the 17th century \citep{Marsden2005}. The probability of such a comet coinciding with astrometric observations of a target star is vanishingly small, and their extreme brightness would make them easily identifiable and excludable from stellar datasets. Therefore, sungrazing comets have negligible practical impact on stellar astrometric jitter studies.

\subsubsection{(Exo-)Zodiacal dust}

Zodiacal light originates from interplanetary dust reflecting stellar radiation within our solar system. Exozodiacal light represents the same phenomenon in other stellar systems. This dust is difficult to observe in detail, and a recent NASA Study Analysis Group outlines several knowledge gaps surrounding exozodiacal dust \citep{SAG23}. When it has been studied, it is typically through the lens of direct imaging of exoplanets (see, for example, \cite{Roberge2012}). Zodiacal light brightness is often quantified in "zodi" units -- the dust luminosity relative to the host star. \cite{Hahn2002} used Clementine spacecraft observations to measure integrated visible zodiacal light at magnitude $-8.5$, corresponding to $L_{dust}/L_\odot \approx 3\times10^{-8}$. Others tend to obtain values around $10^{-7}$ (see \citep{Roberge2012} and the references therein).
If exozodiacal dust were isotropically distributed, its impact on photocentroid measurements would be minimal. While this approximation holds to first order, significant structure can exist. \cite{Hahn2002} found North-South and East-West anisotropies of around 10\% and other smaller scale `clumps' can be linked to trojan dust groups \citep{Stark2011,Montesinos2020} and mean motion resonance \citep{Shannon2015}.

Large-scale anisotropies likely result from collisional processes and persist for tens of thousands of years \citep{Grn1985}.  Trojan dust has even longer lifetimes of $10^5 - 10^6$ years \citep{Liu2018}. Since these timescales far exceed typical astrometric mission durations (a few years), such structures should contribute persistent systematic offsets rather than time-variable jitter.

\begin{figure}
\centering
\includegraphics[width=1\linewidth]{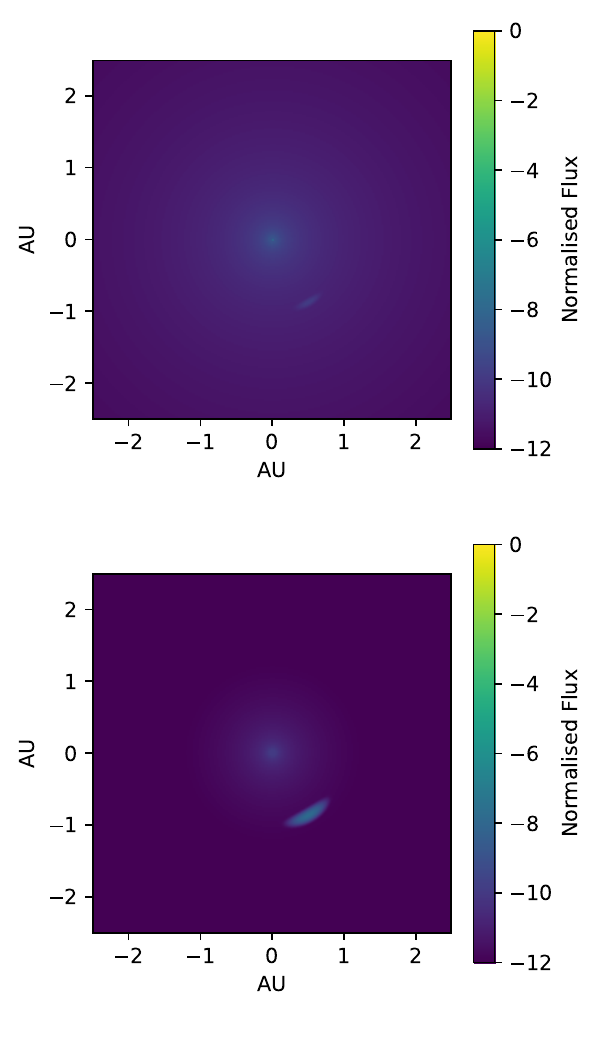}
\caption{The two toy models of trojan dust cloud over-densities. The upper panel is model 1, with a 10x overdensity compared to the same radius background zodiacal dust. The lower panel is model 2, where 90\% of the dust is contained within the trojan dust cloud.}
\label{fig:toymodel}
\end{figure}

To quantify small-scale structure impacts, we developed a toy model using a 2D grid spanning 5 AU centered on a 1-pixel star. We added background zodiacal dust following \cite{Hahn2002}'s density profile $\sigma(r) \propto r^{1.45}$. We then added a gaussian arc overdensity at 1AU to simulate gravitational trapping induced by a planet.  We normalized total dust flux to one zodi ($1e^{-7}$) and test two scenarios: (1) where the overdensity was 10x higher than the background zodiacal dust at the same radius, and (2) where 90\% of the dust was in the overdensity (See Figure \ref{fig:toymodel}). Both models had a FWHM of 0.1 AU, and an azimuthal extent of 30 degrees.

For a solar-level zodi, the results were negligible - $0.5$ nano as pc for model 1, and $0.09\mu$as pc for model 2. \cite{Ertel2020} conducted a survey of Solar-like stars and found that the median zodi level was 3 times higher than the solar value, and at the 2 sigma upper limit was 27 times higher. We took the 27x zodi level and applied it to our toy model. Model (1) had a photocentroid offset of $0.015\mu$ as pc, and model (2) had an offset of $2.42\mu$ as pc. While data surrounding the small scale structure dust densities is sparse, it is unlikely that 90\% of exozodiacal dust is trapped within a trojan dust cloud. Additionally, both trojan and mean motion resonant dust clouds co-rotate with their host planets (the latter harmonically), generating systematic photocentroid offsets rather than time-variable jitter. While such systematic effects could potentially be modelled and corrected, they may create degeneracies with planetary mass estimates during signal extraction. Although our toy model analysis provides initial insights, further work is required to fully assess exozodiacal dust impacts on precision astrometric missions.

\section{ Results}
\label{sec:results}

\subsection{Quantifying Astrometric Jitter}
We measured the astrometric jitter of the Sun across multiple wavelengths over approximately one solar cycle, finding wavelength-dependent variability that ranges from \update{$\sim$0.34 $\mu$as pc} for quiet red wavelengths to $\sim$7 $\mu$as pc for active Ca~{\sc ii} lines. Our results demonstrate that solar astrometric jitter exhibits both wavelength dependence---with bluer and active wavelengths showing significantly larger deviations---and strong temporal variability spanning an order of magnitude across the solar cycle (see \update{Figure} \ref{fig:90dayrolling}). We contrast our results with the literature to measure the astrometric jitter of the Sun in Table \ref{MainTable}, with our measurements displayed in Figures \ref{fig_KDE} and \ref{fig:90dayrolling}. We find mean astrometric jitter values of \update{0.342, 0.532, and 6.918} $\mu$as pc for the red ( \update{607} nm), blue (409 nm) and Ca~{\sc ii} (393 nm) lines respectively, with maximum deviations of \update{5.232, 7.762, and 43.411} $\mu$as pc. Analysis of 90-day periods shows jitter ranges of \update{0.058--1.294} $\mu$as pc (red),  \update{0.082--1.971} $\mu$as pc (blue), and \update{0.647--10.263} $\mu$as pc (Ca~{\sc ii}). We also find that jitter in the equatorial direction tends to be \(\sim10\)\% higher than jitter in the meridional direction (see \update{Figure} \ref{fig_KDE}).

Our results suggest that previous studies underestimate the range of solar astrometric jitter across activity cycles: quiet periods show lower jitter  \update{($\sim0.06-0.08$~$\mu$as~pc)}than previously estimated ($\sim 0.2$~$\mu$as~pc), while active periods show higher jitter \update{($\sim 1.3-2.0$~$\mu$as~pc)} than the \update{$\sim1.1-1.3\mu\text{as pc}$} found in earlier work\update{, depending on waveband and study} (see Table \ref{MainTable} \update{and Section \ref{sec:time_window}} for details). We note that no previous similar work exists relating to active wavelengths, as previous work is focused on continuum observations (see the works listed in Table \ref{MainTable}). As such, this comparison relates only to quiet wavelengths. We further note that the wavelengths investigated are not consistent across studies, so this is not a one-to-one comparison. Yet, this work and previous works both measure or calculate the photometric jitter of quiet wavelengths, which should have similar jitter properties and---with the exception of active lines---should change smoothly as a function of wavelength. Overall, our measurement of the typical jitter levels are usually lower than other work. We emphasise that the observation period and low-vs-high activity definitions vary, which may partially explain these differences. We do note that ours is the only study that directly measures the photometric centroid, while others indirectly infer this measurement. Details of these differences can be found in the discussion. 

 We find that the astrometric jitter varies with wavelength, with bluer wavelengths exhibiting both larger maximum deviations, and a higher standard deviation. This effect is mostly due to the larger relative drop in flux from the regular solar blackbody, and the cooler blackbody of Sunspots in bluer wavelengths. Additionally, we find that active wavelengths - such as the Calcium-{\sc ii} line at 393nm - exhibit jitter an order of magnitude larger than other wavelengths (also see Fig \ref{fig_KDE}). This is due to the enhanced visibility of solar activity, particularly of faculae, networks, and flares. We also find a temporal dependence across all wavelengths, with the magnitude of the jitter changing by an order of magnitude across the solar cycle, as shown in Figure \ref{fig:90dayrolling}. We also note significant jitter variations of up to 0.5 dex on short timescales. Maximum variability \update{of the jitter} occurs at solar maximum for quiet wavelengths and between solar minimum and maxima for active wavelengths.  This has implications for astrometric missions targeting stars other than the Sun. Notably, target suitability may need to be periodically reassessed as jitter can change dramatically on both short and long terms.

\subsection{Correlations between Astrometric Jitter and TSI}

\update{Using the dataset we constructed in Section \ref{sec:correlations}, we first quantify the instantaneous relationship between TSI and photocentre displacement by computing the Pearson correlation \citep{Pearson1895} between the daily TSI and radial astrometric offsets. The resulting coefficients, $r = -0.13$ (blue) and $r = -0.16$ (red), correspond to an $r^2 \lesssim 0.03$ indicating that TSI variations explain less than $3\%$ of the variance in instantaneous astrometric offset. This is expected: many different surface configurations can produce the same photometric deficit (or excess), which need not yield the same photocentre deflection.} 

\update{To investigate whether the variability amplitudes of the two signals are related, we computed 30-day rolling standard deviations of the photocentre displacement and of the TSI, then correlate the paired 30-day values across the entire PSPT timespan. This produces apparently strong positive correlations of $r = 0.835^{+0.016}_{-0.018}$ (blue) and $r = 0.801^{+0.019}_{-0.021}$ (red), with 95\% confidence intervals estimated from 5,000 bootstrap replicates. Thus, when considered over an entire activity cycle, periods of enhanced TSI variability tend to coincide with enhanced astrometric variability. This is sensible, as for the Sun, enhanced TSI variability arises from the appearance of active regions whose brightness contrasts also produce photocentre displacements. Periods of high magnetic activity drive both signals simultaneously, even though their instantaneous values are only weakly correlated.} 

\begin{figure*}
\centering
\includegraphics[width=0.9\linewidth]{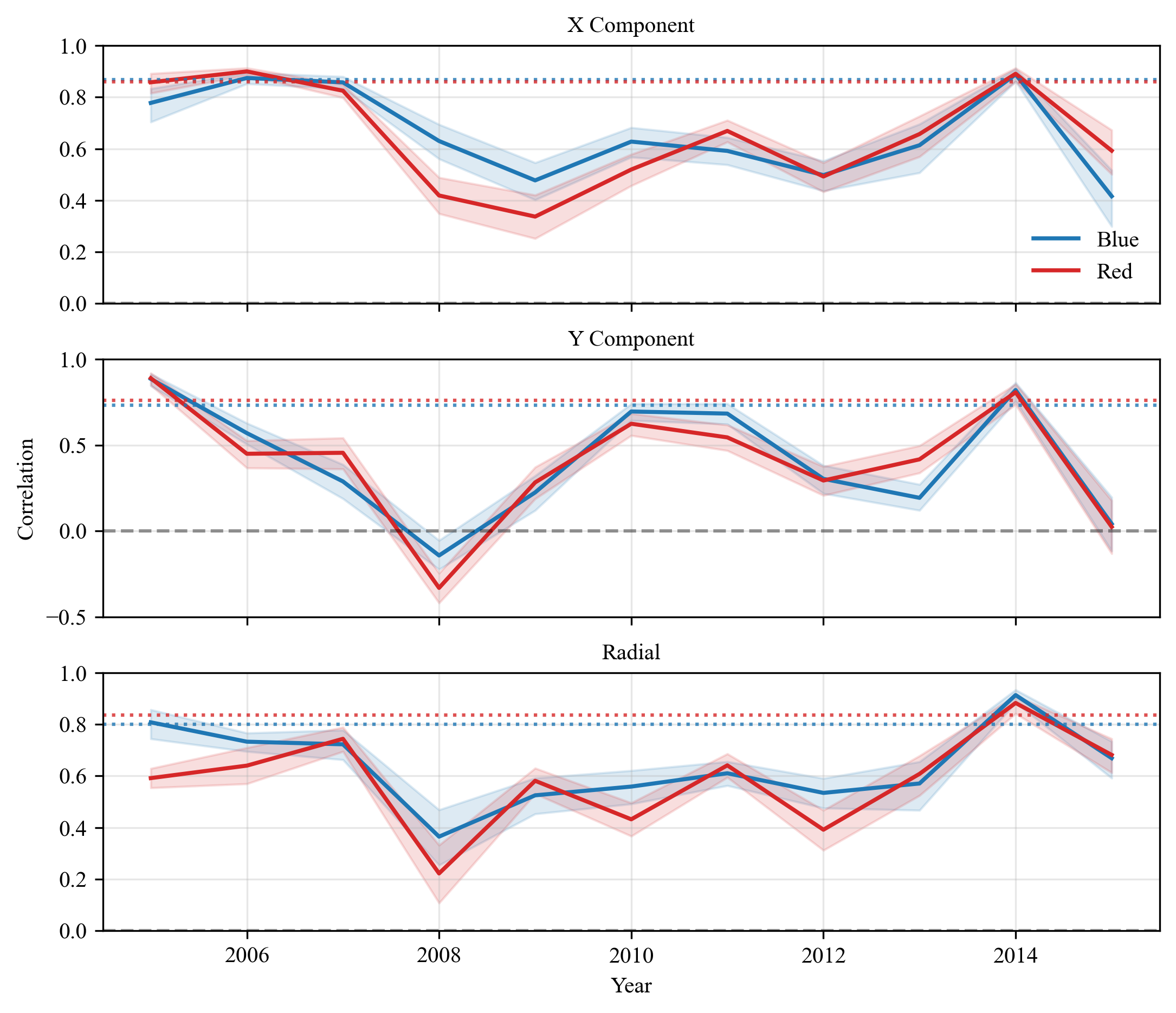}
\caption{\update{Calendar-year correlations between solar irradiance variability and photocentre jitter, by component and waveband. Solid lines represent the per-year point measurements, with the shaded regions showing the 95\% confidence intervals, estimated via bootstrapping. Coloured horizontal dotted lines show the corresponding pooled correlation over the full data range. The grey dashed line marks zero correlation. The first and last years (2005 \& 2015) do not have a full 12 months of data. The magnitude of the X and Radial components roughly track the solar cycle, with generally stronger correlations near solar maxima and weaker correlations during solar minima.}}
\label{fig:peryearcorr}
\end{figure*}

\begin{figure*}
\centering
\includegraphics[width=0.9\linewidth]{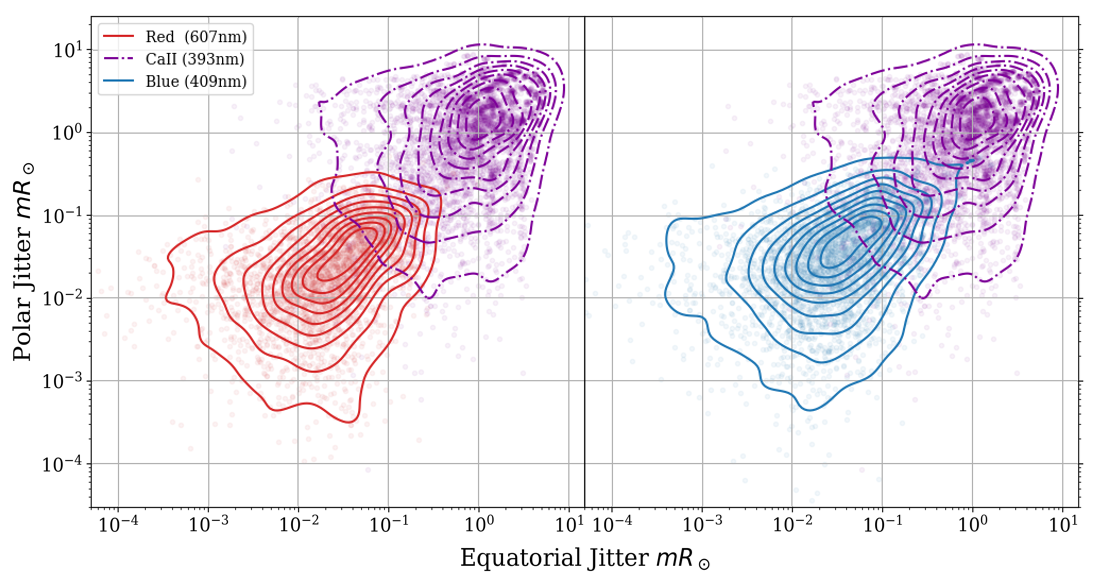}
\caption{Kernel Density Estimates (KDEs) of the astrometric jitter of the Sun aligned in the equatorial frame. The entirety of the PSPT dataset is plotted, demonstrating how the astrometric jitter varies over several orders of magnitude, and is heavily wavelength-dependent. This demonstrates the need for careful waveband selection when designing astrometric telescopes, as including lines associated with stellar activity can lead to a significant increase in observed astrometric jitter.} 
\label{fig_KDE}
\end{figure*}

\begin{figure*}
\centering
\includegraphics[width=0.9\linewidth]{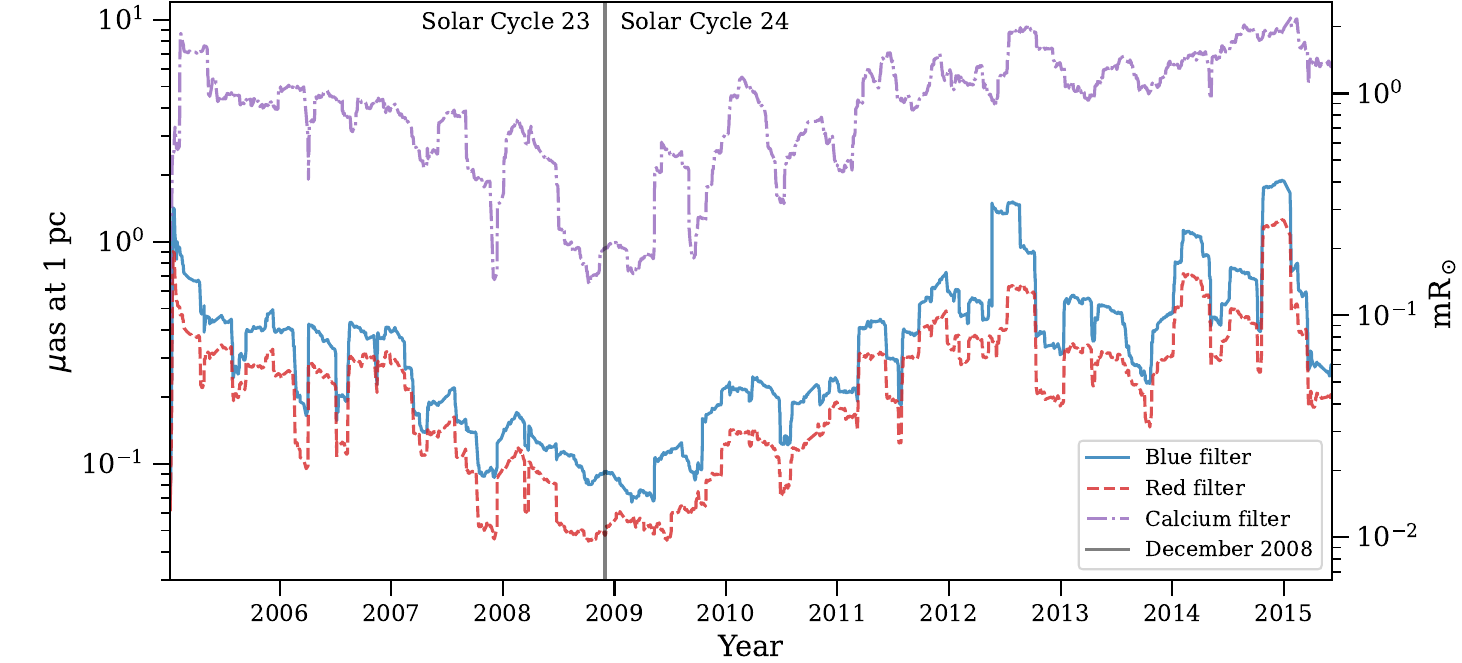}
\caption{The 90-day rolling average of the standard deviation of the absolute photometric deflection (i.e. astrometric signal) of the Sun in three wavelengths, spanning from 2005-2015. The rapid changes in variation indicates that short term jitter trends are heavily impacted by quasi-random spot clusters appearing. The presence of the Solar cycle is evident in this data. Notably, the active wavelength of Calcium-{\sc ii} typically has jitter an order of magnitude larger than quiet wavelengths, such as the red and blue filters. Even in the `quiet` red and blue wavelengths, the jitter varies by an order of magnitude across the Solar cycle. If possible, astrometric missions should aim to observe targets during a period of activity minima in order to maximise the possibility of exoplanet detection.}
\label{fig:90dayrolling}
\end{figure*}

\update{To assess whether this cycle-averaged correlation is stationary and holds for any given time period, we compute calendar-year correlations using only the rolling 30-day standard deviation value pairs for each year (Figure \ref{fig:peryearcorr}). The X (equatorial) and radial components appear strongly correlated during active phases of the solar cycle -- typically $r \gtrsim 0.8$ -- but this weakens during solar minimum, often $0.2 \lesssim r \lesssim 0.7$. The Y (Meridional) component presents greater variability, less tied to the phase of the solar cycle, and briefly becomes negative around the 2008 minimum.}

\update{This behaviour, strong overall correlation, with typically weaker correlations over shorter timescales, is an example of temporal aggregation bias (which is related to Simpson's paradox). Consequently, while photometric variability can have some use as an indicator of astrometric jitter in an average sense, the relationship should not be assumed to provide a stable or universal predictor of centroid for all epochs or activity phases. We caution against inferring astrometric stability from photometric quietness alone. What this does indicate is that periods of elevated photometric variability may coincide with elevated astrometric jitter for Sun-like stars with similar surface properties. }

\update{When measuring the strength of correlations, the choice of window size for the standard deviation does have an effect. This effect is demonstrated in Figure \ref{fig:windowsize}. For small correlation windows, the correlation is weak. The correlation grows quickly up to around 30-days, before starting to plateau for longer windows. We selected 30-days for our period as it is close to the solar rotation period. Windows substantially longer than this offer diminishing returns, as they average over multiple rotation periods, removing the connection to individual active regions. The one exception to this plateau is in the Y (meridional) component of the red waveband. }


\begin{landscape}
\begin{table*}
\raggedright 
\hspace*{-2.5in} 
\rowcolors{2}{gray!25}{white} 
\begin{tabular}{|p{4cm}|p{2.5cm}|p{1.8cm}|c|c|c|c|c|}
\hline
\rowcolor{gray!50}
\textbf{Name} & \textbf{Method} & \textbf{Wavelength(s) (nm)} & \textbf{Typical Dispersion} (\(\mu\)as pc) & \textbf{Max Deviation*} (\(\mu\)as pc) & \textbf{Low Activity} (\(\mu\)as pc) & \textbf{High Activity} (\(\mu\)as pc) \\ \hline
\cite{Hatzes2002} & Multi-Spot Model & - & 2.82 (a) & 16.15 (b) & - & - \\ \hline
\cite{Bastian2004} & - & - & - & 25 & - & - \\ \hline
\cite{Reffert} & Single Spot Model & 2000-2400 & - & 10 & - & - \\ \hline
\cite{Eriksson2007} & Multi-Spot Model & - & - & - & $<$5 (c) & $>$10 (c) \\ \hline
\cite{Lanza2008} & Single Spot Model & - & 2.60 (a) & 14.6 (b) & - & - \\ \hline
\cite{Catanzarite2008} & Multi-Spot Model & \update{Bolometric} & 0.72 (c) & - & - & - \\ \hline
\cite{Makarov2009} & Single Spot Model & - & 1.1 (d) & 1.5 & - & - \\ \hline
\cite{Makarov2010} & Magnetogram Reconstruction & Bolometric & 0.65 (d)(k) & 2.6 (e)(k) & - & 0.91 (e)(k) \\ \hline
\cite{Lagrange2011} & Multi-Spot Model \update{/ Magnetogram Reconstruction} & Bolometric & 0.86 (d) & 20 (f) & 0.22 (d) & 1.08 (d) \\ \hline
\cite{Morris2018} & Multi-Spot Model & 330-1050 & 0.13 (d)(g) & - & - & - \\ \hline
\update{\cite{Meunier2020}} & Simulation & - & 1.06(d) & - & - & 2.66 (d)(h) \\ \hline
\cite{Shapiro2021} & Single Spot Model & 330-1050 & 3.5(i) & 15 & - & 9 (j) \\ \hline
\cite{Shapiro2021} (cont.) &  Magnetogram Reconstruction & \update{330-1050} & - & 6.95 & 0.19 (c) & 1.12(c) \\ \hline
\cite{Shapiro2021} (cont.) &  Magnetogram Reconstruction & \update{330-680} & - & 8.06 & 0.22 (c) & 1.30(c) \\ \hline
\cite{Shapiro2021} (cont.) &  Magnetogram Reconstruction & \update{680-1050} & - & 5.91 & 0.16 (c) & 0.95(c) \\ \hline

\cite{KaplanLipkin2022} & Multi-Spot Model / Magnetogram Reconstruction & 330-680 & 2.33 & - & - & - \\ \hline
This work (Red) & Photometry & 607 &  \update{0.342} (k)(l) & \update{5.232} &  \update{0.058}(k) &  \update{1.294}(k) \\ \hline
This work (Blue) & Photometry & 409 &  \update{0.532} (k)(l) &  \update{7.762} &  \update{0.082}(k) &  \update{1.971}(k) \\ \hline
This work (Ca~{\sc ii}) & Photometry & 393 &  \update{6.918}(k)(l)&  \update{43.411} &   \update{0.647}(k) &  \update{10.263}(k) \\ \hline
\end{tabular}
\caption{A comparison between the absolute (radial) astrometric jitter values in literature. Note that all values have been reported here in micro-arcseconds assuming the star is observed from 1 pc\protect\footnote{Note that these values can easily be converted to an absolute distance, as \(1\mu as = 1\mu \text{AU}\) at 1pc.}, and have been translated as necessary from the units they were originally reported in. Missing values originate from unreported values in the original papers.}
\begin{itemize}
    \item * Unless stated otherwise, half peak to peak
    \item (a) max displacement (half peak-to-peak) for a typical sunspot filling factor of 0.03\% (See Morris et al 2018)
    \item (b) max displacement (half peak-to-peak) for a near max observed sunspot filling factor of 0.2\% (See Lanza et al 2007)
    \item (c) Root Mean Square (RMS)
    \item (d) Paper originally reported X (equatorial) and Y (polar) astrometric deflections separately, but have been added here in quadrature
    \item (e) Result is for X (equatorial) direction only
    \item (f) Not claimed as largest deflection, but claims deflections are typically below this level
    \item (g) Median absolute deviation
    \item (h) Value is for higher spot contrasts, not high activity
    \item (i) Mean max displacement for typical sun filling factor
    \item (j) Mean max displacement for sun during active phase
    \item (k) Standard Deviation
    \item (l) The per image measurement uncertainty is \(10^{-4} \mu\text{as pc}\)
\end{itemize}
\label{MainTable}
\end{table*}
\end{landscape}

\section{Discussion}
\label{sec:discussion}
\subsection{This work}

Our work takes directly measured photometric data of the Sun to calculate the activity induced astrometric deflections, treating the Sun as a star, observed from a distance of 1 parsec. We find significant wavelength dependence for this astrometric signal. At the red continuum (\(\lambda = 607.2\text{nm}\)), we find a typical jitter of \update{\(0.342\mu\text{as pc}\)}, with this dispersion measure ranging between \update{\(0.058\mu \text{as pc}\)} during the solar minimum between cycle 23 and 24 to  \update{\(1.294\mu\text{as pc}\)} during the maximum of cycle 24. This temporal variation is greater than one order of magnitude and is clearly visible in Figure \ref{fig:90dayrolling}. Over the solar cycle the strong dependence of astrometric jitter on overall stellar activity is apparent\update{,} \update{emphasising} the importance of the activity cycle on target selection. \update{Using Equation \ref{eq_astrometry_signal}, an Earth-mass planet at 1\,AU around a solar-mass star at 1\,pc produces an astrometric signal of $\sim 3\,\mu$as, yielding a signal-to-jitter ratio that varies between $\sim 2$--$50$ across the activity cycle, with a typical value of $\sim 9$. A Mars-mass planet at 1\,AU would produce an astrometric deflection of $\sim 0.32\,\mu$as, comparable to our typical red jitter, while Mars at its actual orbital separation of 1.52\,AU would produce $\sim 0.49\,\mu$as. This places the activity-imposed detection floor for a planet orbiting at 1\,AU from a solar analogue at 1 pc observed in the red continuum between Mars-mass and Earth-mass planets, with the exact limit depending on the phase of the stellar activity cycle.}

The astrometric jitter shows strong wavelength dependence (Figures \ref{fig_KDE} and \ref{fig:90dayrolling}). The activity-sensitive Ca~{\sc ii} line (\(\lambda = 393.4\)nm) exhibits jitter an order of magnitude larger than the red and blue continuum wavelengths throughout the solar cycle. The blue wavelength (\(\lambda = 409.3\)nm) consistently shows larger astrometric signals than the red (\(\lambda = 607.2\)nm), demonstrating the importance of passband selection for astrometric missions targeting microarcsecond precision. As highlighted by \cite{KaplanLipkin2022}, the relative contrast between the quiet sun, faculae, and sunspots are higher towards the peak of the blackbody radiation distribution than the (redder) tail. This leads to larger photometric deficiencies for spots (and larger excesses for bright features), \update{producing} a larger deflection of the photometric centroids and hence a larger jitter. \update{The PSPT narrowband measurements sample the continuum at very narrow set of wavelengths, but the underlying physics of reduced spot and faculae contrast at longer wavelengths applies broadly. }Astrometric missions aimed at finding exoplanets should  \update{favour} passbands towards the tail of the blackbody distribution to maximise the detectability of any exoplanets, \update{while balancing flux requirements}.

Our analysis is limited to observations of a single star: the Sun. While these results represent the best available data for estimating astrometric jitter in Sun-like stars, extrapolation to other systems requires caution regarding the applicability of this data. For example, the Sun is photometrically quiet compared to other solar-like stars \citep{Reinhold2020}. We further note that other stars will have different metallicities, ages, and activity patterns that will likely alter the jitter profiles (see \cite{Karoff2018}). \cite{Sowmya2022p3} investigate the impact of rotation rate on astrometric jitter --- with temperature and metallicity fixed at solar values -- and find that faster rotators have higher magnetically induced astrometric jitter, as surface activity scales with rotation rate. Another limitation \update{is that} our observations  are constrained to a single viewing geometry (Earth's orbital plane), whereas exoplanet surveys will observe stars at random inclinations, potentially altering the observed jitter \update{\citep[See, e.g., ][for further discussion on the impacts of inclination]{Lanza2008,Meunier2020}}. The observed jitter magnitude and anisotropy will depend on viewing inclination due to the latitudinal distribution of stellar activity. As discussed previously, the Sun's equatorial jitter is $\sim$10\% larger than meridional jitter when viewed from the ecliptic plane. This anisotropy arises because spots can appear at any longitude but are latitudinally confined to activity bands that tend to not reach the poles, creating greater variation in the equatorial direction. For pole-on viewing geometries, this anisotropy should disappear and jitter should be equal in both directions. The quasi-periodic nature of stellar activity (tied to rotation) versus the strictly periodic planetary signal can enable frequency domain separation techniques. Care should be taken when choosing targets to ensure the stellar rotation period differs sufficiently from the expected planetary orbital period to avoid signal confusion. 

For missions like TOLIMAN \citep{Tuthill2018} and ARMADA \citep{Gardner2022} --- or any narrow angle astrometry mission --- that employ relative astrometry of binary stars, the measurement strategy fundamentally differs from absolute stellar position monitoring. Rather than tracking a single star against distant reference sources, TOLIMAN measures the separation vector between the two components of $\alpha$ Centauri AB. This differential approach achieves better precision but introduces new complexities. Each stellar component contributes its own activity-induced jitter, and these signals must be disentangled from genuine orbital perturbations caused by planetary companions. For \update{such} missions, an ideal target would be a binary with sufficiently different rotation periods, as these temporal signatures can help separate the two stellar astrometric signals. Additionally, simultaneous photometric monitoring of both stars can track their individual activity cycles, providing independent constraints on the expected jitter contributions that aid in isolating genuine gravitational perturbations from surface activity noise.

\subsection{ Overview of Previous Approaches}
Previous estimations of astrometric jitter - both for the Sun and other stars - show considerable variability, with typical dispersion estimates ranging from 0.13 $\mu$as pc \citep{Morris2018} to 1.06 $\mu$as pc \citep{Meunier2019} for solar-like conditions. These estimates broadly fall into three categories: Estimations using a single spot model (see \cite{Reffert}; \cite{Muterspaugh2006}; \cite{Eriksson2007}; \cite{Lanza2008}; \cite{Makarov2009}; \cite{Shapiro2021}), multi-spot model (see \cite{Hatzes2002}; \cite{Sozzetti2005}; \cite{Eriksson2007}; \cite{Catanzarite2008}; \cite{Lagrange2011}; \cite{Morris2018}), or reconstructions of magnetograms (see \cite{Makarov2010}; \cite{Shapiro2021}). A summary of the results from these papers can be found in Table \ref{MainTable}. Certain models don't fit into this categorisation (such as \cite{Meunier2019}), but still estimate the astrometric jitter primarily based off of the large scale magnetic surface features such as spots and faculae. In addition to these estimations, there are works that calculate the astrometric jitter induced by stellar granulation (\cite{Ludwig2005}; \cite{Bastian2004}). While the contribution from granulation is small for solar-like stars - on the order of \(0.1\mu \text{as pc}\) - the effect scales with the characteristic size of the convection cells and hence can be significant for stars with low surface gravity (\cite{Ludwig2005}). 

Single spot models, while unrealistic, provide upper limits since they concentrate all photometric variability into one feature. These models consistently predict jitter levels of 1--3 $\mu$as pc for typical solar conditions, with extreme cases reaching 10+ $\mu$as pc. For example, \cite{Reffert} finds 10 $\mu$as pc deflections for 0.5\% filling factors, while \cite{Lanza2008} predicts 2.6 $\mu$as pc for typical solar conditions. The range of relationships reported by different groups \citep{Eriksson2007, Makarov2009} reflects the sensitivity of these models to their underlying assumptions about spot properties and stellar parameters.

Multi-spot models represent a more realistic approach to stellar surface modeling, incorporating multiple distributed features rather than a single dominant spot. These models consistently predict jitter levels of 0.7--0.9 $\mu$as pc for typical solar conditions---significantly lower than single-spot upper limits but still higher than our direct measurements. This convergence across different multi-spot approaches, with \cite{Catanzarite2008} reporting 0.72 $\mu$as pc and \cite{Lagrange2011} finding 0.86 $\mu$as pc, suggests these models capture the statistical behaviour of distributed activity better than single-spot \update{models}.

\begin{figure*}
\centering
\includegraphics[width=0.9\linewidth]{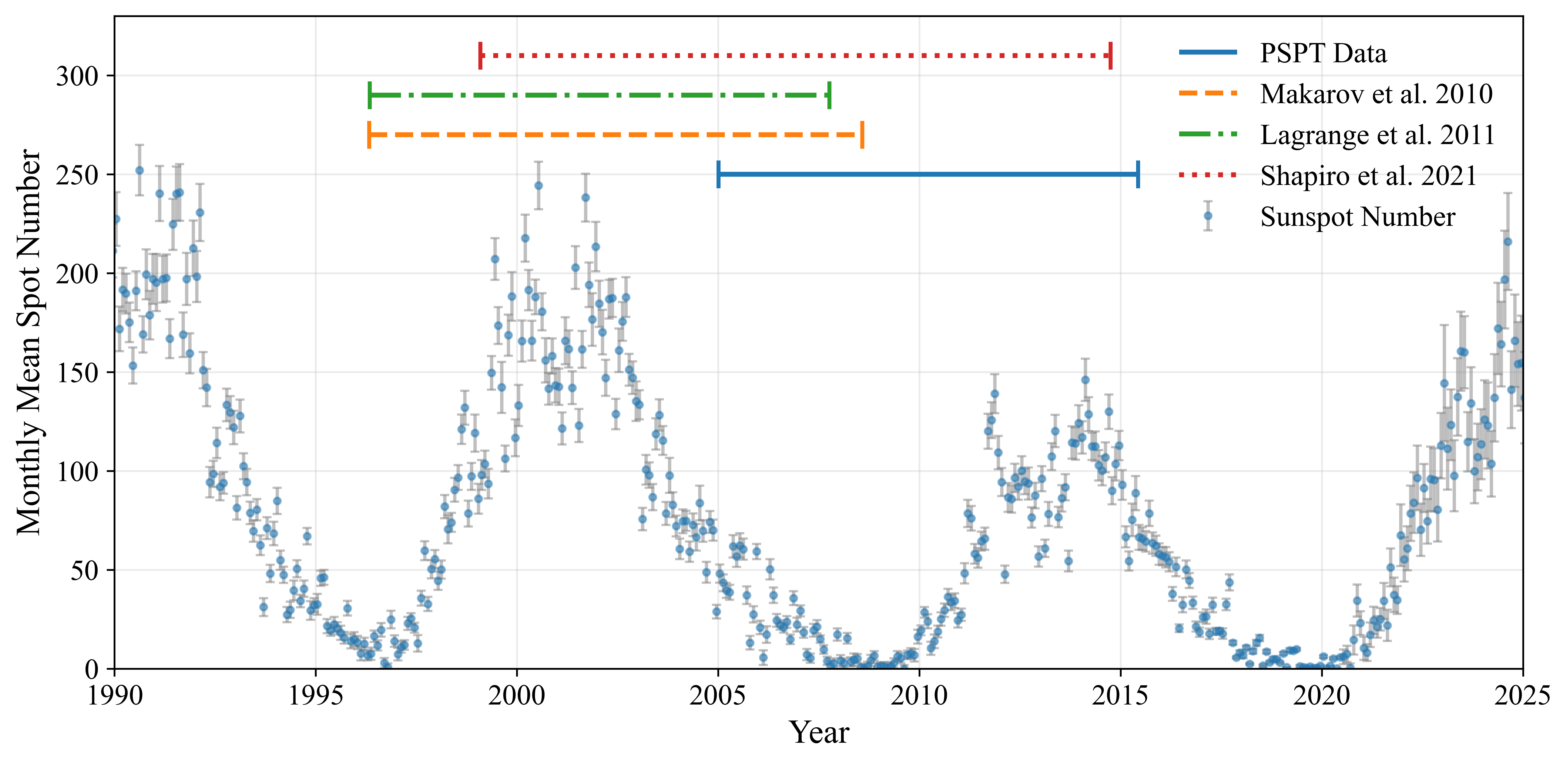}
\caption{Monthly mean International Sunspot Number from the World Data Center SILSO, Royal Observatory of Belgium \citep{SILSO}. Error bars represent the standard deviation among contributing observatories. Horizontal bars indicate the temporal coverage of solar astrometric studies: this work (PSPT data), \citet{Makarov2010}, \citet{Lagrange2011}, and \citet{Shapiro2021}. Despite the cycle that the PSPT data covers being weaker than what other works cover, we get comparable or higher typical astrometric jitters.} 
\label{fig_solarcyc}
\end{figure*}
\begin{figure}
\centering
\includegraphics[width=0.9\linewidth]{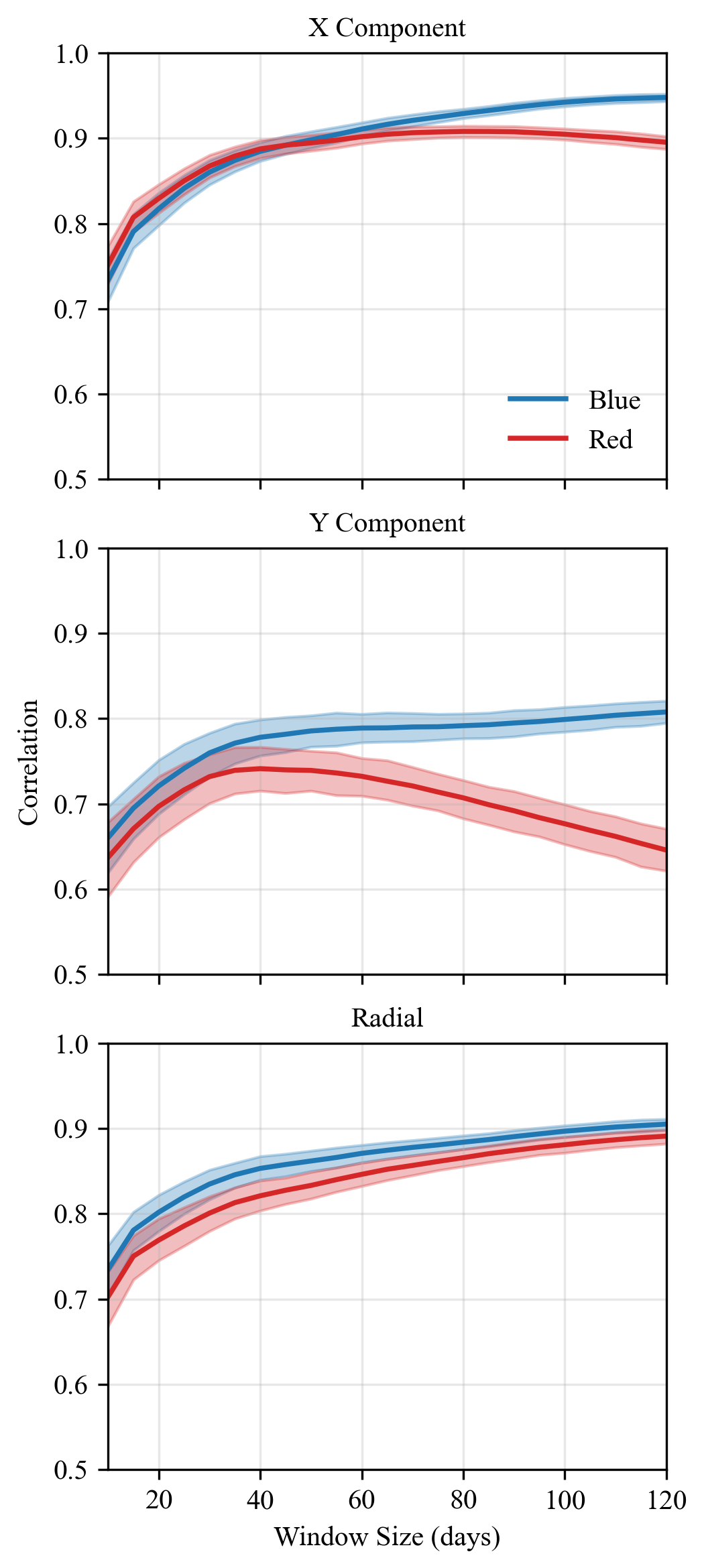}
\caption{ \update{Pearson correlations between the rolling standard deviations of TSI and astrometric jitter as a function of window size. Panels show the X -- equatorial -- component (top), Y -- polar -- component (middle), and radial magnitude R (bottom). Curves are for the blue and red wavebands, with shaded regions being 95\% confidence levels, estimated via bootstrapping. Most display similar shapes: An initial rapid increase of correlation strength with window size up to a size of about 30 days, where a `knee` in the curves leads to a plateauing of the curves. A notable exception is the Y component of the red curve, which shows a decreasing trend after the knee.}}
\label{fig:windowsize}
\end{figure}

\update{A more in-depth approaches use magnetogram data through either intensity reconstruction by classifying pixels into empirically TSI-calibrated brightness classes, or semi-empirical modelling which weights precomputed component intensities by magnetogram-derived fractional coverages of faculae, umbra, penumbra, and quiet Sun. This category, populated by \cite{Makarov2010}, \cite{Lagrange2011}, and \cite{Shapiro2021}, use varying methodologies and data sources, but all incorporate real solar observations spanning substantial portions of solar cycles (see Figure \ref{fig_solarcyc}). \cite{KaplanLipkin2022} investigate wavelength-dependent jitter using SATIRE-S, though their work relies on the \cite{Shapiro2021} time series and is therefore not independent. \cite{Meunier2020} investigate the relationship between stellar activity and astrometric jitter for spectral types F6-K4 at various inclinations, reporting typical jitter of \(1.06-2.66\mu \text{as pc}\) for edge on Sun-like stars (extracted from their Figure 1, assuming G2 spectral type) depending on assumed spot contrast . This is not a direct solar measurement but rather a simulation applicable to G2 stars broadly.}

\update{We more directly focus our comparison on \cite{Makarov2010}, \cite{Lagrange2011}, and \cite{Shapiro2021}, as these are the most directly comparable work.}

\update{\subsection{Comparison with Directly Comparable Studies}}

\update{Figure \ref{fig_solarcyc} shows the mean monthly spot number from 1990 to 2025 with the temporal coverage of solar astrometric studies indicated. The PSPT data spans 2005-2015, covering the decline of Solar Cycle 23 and most of Solar Cycle 24. \cite{Makarov2010} and \cite{Lagrange2011} both cover the majority of Solar Cycle 23, which was more active than Cycle 24 with a peak monthly sunspot number of $\sim250$ vs $\sim 150$. \cite{Shapiro2021} covers most of both cycles. These three studies employ magnetogram-based approaches and report activity-dependent jitter values, making them the most directly comparable to our work.}

\subsubsection{Time window effects and direct comparison}
\label{sec:time_window}
\update{Comparing jitter values across studies requires care, as each study defines low and high activity periods differently, and these definitions can shift reported values by a factor of two or more. We isolate this effect before examining residual differences that cannot be attributed to windowing.}

\update{Our primary low-activity values in Table \ref{MainTable} represent the minimum of a 90-day rolling standard deviation across the full PSPT dataset. This measure is designed to capture the quietest sustained period. In contrast, \cite{Shapiro2021} define their low-activity period as the full calendar year 2008, and \cite{Lagrange2011} use a 9-month window from July 1996 to April 1997 during Cycle 23. \citet{Makarov2010}, on the other hand, use a three-month period similar to us.}

\update{To enable a like-for-like comparison, we computed our jitter over identical time windows where the data overlaps. Only \citet{Shapiro2021}'s data overlaps for a period they consider low activity. For the calendar year 2008, we measure a red-filter standard deviation of $0.074\mu \text{as pc}$ and a blue-filter standard deviation of $0.125 \mu \text{as pc}$. \citet{Shapiro2021} report low-activity values of $0.22\mu\text{as pc}$ (330--680\,nm), $0.16\mu\text{as pc}$ (680--1050\,nm), and $0.19\mu\text{as pc}$ (full 330--1050\,nm range). Our blue filter (409\,nm) falls within their 330--680\,nm band, and our red filter (607\,nm) also falls within this band, though near its red edge. Both of our 2008 values are lower than all three \citet{Shapiro2021} estimates for the same period, by factors of $\sim1.3-3\times$. For the calendar year 2014, the most active year covered by the PSPT dataset, we measure a standard deviation of $0.781\mu \text{as pc}$ (red) and $1.201\mu \text{as pc}$ (blue).}

\update{This window-matched comparison reveals an asymmetry. During low activity (2008), our values fall below all three \citet{Shapiro2021} waveband estimates for the same period. During high activity (2014), our blue measurement ($1.201\mu\text{as pc}$) exceeds the values reported by \citet{Makarov2010} ($0.91\mu \text{as pc}$) and \citet{Lagrange2011} ($1.08\mu \text{as pc}$) by $11$--$32\%$, but falls $\sim 8\%$ below \citet{Shapiro2021}'s 330--680\,nm estimate of $1.30\mu\text{as pc}$ for the same wavelength range. Our red measurement ($0.781\mu\text{as pc}$) falls below \citet{Shapiro2021}'s 680--1050\,nm value of $0.95\mu\text{as pc}$, though this comparison is less direct as our 607\,nm filter sits near the boundary of their two bands. Significantly, these comparisons are made despite Solar Cycle 24 being substantially weaker than Solar Cycle 23 (peak monthly spot number of $\sim150$ vs $\sim250$; see Figure \ref{fig_solarcyc}), the peak of which is covered by all three prior works. If windowing were the sole explanation for all discrepancies, we would expect our Cycle 24 values to fall systematically below their Cycle 23 values. While our year-averaged blue value is indeed slightly below \citet{Shapiro2021}'s wavelength-matched estimate, it exceeds \citet{Makarov2010} and \citet{Lagrange2011} despite the weaker cycle. Furthermore, our 90-day rolling maximum blue jitter of $1.971\mu\text{as pc}$ substantially exceeds all prior high-activity estimates, including \citet{Shapiro2021}'s 330--680\,nm value, indicating that direct photometric measurements capture larger peak excursions during active periods. We explore alternative explanations in forthcoming sections.}

\update{We note that a direct window-matched comparison with \citet{Lagrange2011} and \citet{Makarov2010} is not possible, as their high-activity periods were measured during Solar Cycle 23, outside the range of the PSPT data. However, their high-activity values of $1.08\,\mu\text{as\,pc}$ and $0.91\,\mu\text{as\,pc}$ were measured during a notably stronger cycle than ours (peak monthly spot number $\sim250$ vs $\sim150$), yet our blue-filter value of $1.201\,\mu\text{as\,pc}$ from the weaker Cycle 24 still exceeds both. Our overall red and blue dispersions ($0.342$ and $0.532\,\mu\text{as\,pc}$) are both lower than the typical values reported by \citet{Makarov2010} ($0.65\,\mu\text{as\,pc}$) and \citet{Lagrange2011} ($0.86\,\mu\text{as\,pc}$), which may be partially attributable to our narrowband filters excluding active-line contributions (see Section~\ref{sec:activeline_and_wavelength}). This combination of lower overall dispersion with comparable or higher active-period values, despite a weaker solar cycle, suggests that direct photometric measurements may capture a wider dynamic range of jitter than reconstruction-based approaches.}

\subsubsection{Wavelength coverage and active line contributions}
\label{sec:activeline_and_wavelength}
\update{Our measured astrometric jitter is \(0.342\), \(0.532\), and \(6.918 \mu \text{as pc}\) for the Red, Blue, and Ca~{\sc ii} wavebands, respectively. Our Red and Blue values are notably lower than the \(0.65\) and \(0.86  \mu \text{as pc}\) reported by \cite{Makarov2010} and \cite{Lagrange2011}, respectively. One possible explanation is wavelength coverage as the latter results represent bolometric or broadband wavelengths, while our measurements use narrow ($<1\,\text{nm}$) filters. Broader wavebands necessarily include flux from magnetically active spectral lines, which could elevate the measured jitter.}

\update{To determine whether wavelength coverage explains this differences, we analysed the fractional contribution of absorption lines to solar flux in our wavelength region. We used high-resolution ($R_s > 100,000$) HARPS solar spectra \citep[SNR > 1000,][]{Dall2006} which spans 378-691nm with \(\sim 0.01\) \r{A} sampling. We extracted a pseudo-continuum by applying a maximum filter with a window of $7\,\text{nm}$ for $\lambda < 400\,\text{nm}$ (where broad features span several nanometres) and $4\,\text{nm}$ at longer wavelengths. Wavelengths with flux below 0.9 of the pseudo-continuum were classified as absorption lines. This threshold is heuristic, this excludes the core of the lines and most of the wings. Varying the threshold between $0.85$ and $0.95$ changes the estimated line flux from approximately $30\%$ to $57\%$, reflecting the inherent ambiguity in defining where lines end and the continuum begins in a crowded spectrum. At our adopted threshold, approximately $39.5\%$ of the flux comes from the absorption lines. Following \cite{Dumusque2018}, who finds that that approximately half of spectral lines are activity sensitive for radial velocity, we estimate that roughly $15-30\%$ of solar flux comes from activity sensitive lines. We take $20\%$ as a representative value.}

\update{Using this estimate, we can bracket the expected jitter for a broadband observation. If the activity-sensitive lines exhibit variations as strong as our Ca~{\sc ii} measurement (an upper bound), a linear combination of 80\% continuum ($0.342\mu \text{as pc}$) and 20\% Ca~{\sc ii}-like ($6.918 \mu \text{as pc}$) contribution yields $\sim1.7\mu \text{as pc}$ (correlated) or $\sim1.4\mu \text{as pc}$ (uncorrelated, added in quadrature). Both exceed the $0.65-0.86 \mu \text{as pc}$ reported by \citet{Makarov2010} and \cite{Lagrange2011}. If instead the activity-sensitive lines exhibit half the variance of Ca~{\sc ii}, the estimates fall to $\sim 1\mu \text{as pc}$ (correlated) or $\sim 0.76\mu \text{as pc}$ (uncorrelated), closer to previous works.}

\update{The above estimates should be treated cautiously. The Ca~{\sc ii} K line is a strong chromospheric line that responds directly to stellar magnetic fields  \citep[see, e.g.,][]{Hall2008}, while much of the forest of weak metallic lines forms in the photosphere where gas pressure dominates and different mechanisms, such as convective blueshift inhibition apply \citep{Cretignier2020}. Many smaller absorption lines need not behave like  Ca~{\sc ii}, and the assumption that active-line radial velocity behaviour translates to astrometric behaviour is speculative. To the best of our knowledge, there have been no comprehensive studies of the impact of individual active lines on astrometric jitter.}

\update{Wavelength coverage can account for why our typical jitter is lower than broadband estimates as our narrowband filters exclude active line flux that would elevate the measured dispersion. However, this effect should be stronger during active periods, when magnetically sensitive lines exhibit their largest variations. Our Ca~{\sc ii} demonstrates this as the jitter tracks the solar cycle in phase with the continuum, differing only in amplitude (Figure \ref{fig:90dayrolling}). Broadband inclusion of such lines should amplify the activity dependence of the jitter, predicting that our narrowband active period values would fall even further below broadband estimates. Instead we observe the opposite. The asymmetry of lower quiet period values but higher active period values cannot be fully attributed to wavelength coverage and likely reflects methodological differences in how the reconstructions recover jitter amplitude. We examine these differences in Section \ref{sec:method_diffs}.}


\subsubsection{Ruling out spot cancellation}
\label{sec:spot_cancellation}

\update{One possible explanation for our higher active-period jitter relative to the reconstruction studies is spot cancellation as during periods of higher activity, more spots are present simultaneously and their individual photocentre deflections could partially cancel, suppressing net jitter. To test this, we simulated stellar surfaces with spot counts ranging from one to 900 under three distribution schemes: (1) uniformly-sized spots placed randomly in $\ell$ and $\cos(b)$, (2) uniformly-sized spots evenly distributed via a Fibonacci spiral, and (3) a realistic size distribution following the solar butterfly pattern described in Section \ref{sec:fulldisk}.}

\begin{figure}
\centering
\includegraphics[width=1\linewidth]{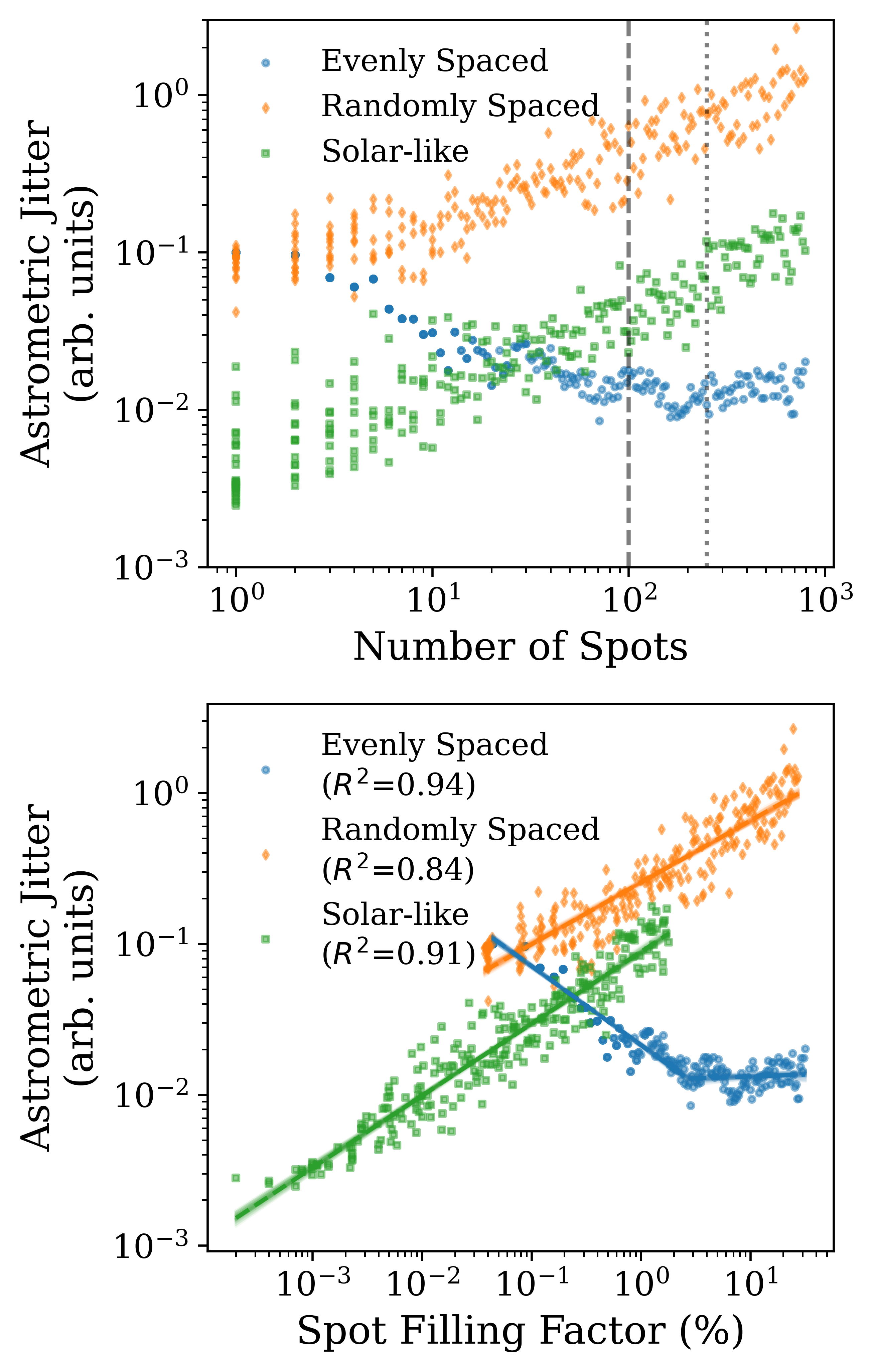}
\caption{The relationship between number of spots and astrometric jitter (upper) and the spot filling factor and astrometric jitter (lower) for three different spot distribution schema. The blue circles represent the evenly spaced schema, where spots of uniform size are spaced according to a fibonacci spiral, for maximum astrometric deflection cancellation. The orange diamonds represent the randomly spaced schema, where spots of uniform size are placed randomly in longitude and the cosine of latitude. The green squares represent the solar like schema as described in Section \ref{sec:fulldisk}. \update{Overlaid in the lower plot are lines fitted to the data in log-log space, corresponding to a power law for the orange diamonds and green squares, and a broken power law for the blue circles. In log-log space, the slope for the randomly spaced spots is $0.410^{+0.011}_{-0.011}$, the slope for the solar-like spot distribution is $0.477^{+0.010}_{-0.009}$. For the evenly spaced spots, the slope before the break is $-0.523^{+0.012}_{-0.012}$, the slope after the break is $0.025^{+0.028}_{-0.028}$, and the break occurs at a spot filling factor percentage of $2.67^{+0.23}_{-0.27}\%$. The uncertainties reported above are the 68\% credible interval.}}
\label{fig:paints}
\end{figure}

\update{For the random and solar-like distribution (schemes 1 and 3), jitter increases as a power law with spot filling factor, with log-log slopes of $0.410^{+0.011}_{-0.011}$ and $0.477^{+0.010}_{-0.009}$ respectively in log-log space (68\% credible intervals). No cancellation is observed at any tested filling factor for these schemes. Only the Fibonnaci distribution (schema 2) exhibits cancellation, following a broken power law with an initial slope of $-0.523^{+0.012}_{-0.012}$ that flattens to a slope consistent with zero ($0.025^{+0.028}_{-0.028}$) above a filling factor of $2.67^{+0.23}_{-0.27}\%$ (Figure \ref{fig:paints}). However, this configuration is deliberately conservative. The evenly spaced spots maximise cancellation by distributing intensity deficits near-optimally around the disc centre. In reality, solar active regions emerge in spatially clustered nests at preferred longitudes \citep{Castenmiller1986}. Clustered spots produce deflection vectors that point in similar directions and therefore mostly add rather than cancel, unlike the evenly distributed case where vectors span all directions. Even for this efficient configuration, the cancellation plateau only begins at a filling factor of about $2.7\%$, roughly nine times the $\sim0.3\%$ typical of the active Sun. For the more realistic distributions, any such plateau would lie well above this threshold. We therefore conclude that spot cancellation does not explain the discrepancy between our results and those of \citet{Makarov2010}, \citet{Lagrange2011} or \citet{Shapiro2021}.}

\subsubsection{Methodological origins of differences}
\label{sec:method_diffs}
\update{Having ruled out spot cancellation and established that wavelength coverage can account for some but not all of the discrepancy, we now examine the methodological choices of each reconstruction study.}

\update{\cite{Makarov2010} report a typical astrometric jitter of \(0.65\mu \text{as pc}\), rising to \(0.91\mu \text{as pc}\) during high activity periods. They use images reconstructed by \cite{Ulrich2009} from Mount Wilson Observatory data, combining magnetograms and two narrowband photometric images (FeI \(5250.2\)\r{A} and CrII \(5237.3\)\r{A}). The photometric images are ratioed, which removes limb darkening, and pixels in the reconstructed images are classified into 18 classes. A multiple linear regression determines coefficients for each class that best reproduced total solar irradiance (TSI) as measured by VIRGO/SOHO achieving a correlation of \(r=0.9625\). While this demonstrates accurate recovery of bolometric intensity, the resulting images are not bolometric intensity maps. As \cite{Ulrich2009} states each pixel represents local contribution to the change in TSI, not the local photometric contrast. This distinction leads to image artifacts, such as bright ringing around active regions that can reverse the sign of the feature. A dark region in a true bolometric map may appear bright in the reconstruction (see Figure 8 of \cite{Ulrich2009}). Such sign reversals directly affect the computed photocentre, causing features which should repel the photocentre to instead attract it. More generally, errors that cancel when integrating the zeroth moment (TSI) need not cancel when integrating the first moment (photocentre). The \citet{Makarov2010} high-activity jitter is smaller than both our red and blue values, while typical jitter is higher, a pattern that is not easily explained by any single systematic effect such as the absence of limb darkening. This suggests that TSI-calibrated reconstruction may not fully capture the solar astrometric jitter.}

\update{\cite{Lagrange2011} report a typical astrometric jitter of \(0.86 \mu \text{as pc}\), a jitter of  \(1.08 \mu \text{as pc}\) during periods of high activity and \(0.22 \mu \text{as pc}\) during periods of low activity. They use spot positions from the USAF/NOAA catalogue and plage locations extracted from MDI/SOHO magnetograms placing circular spots at the catalogued positions with areas converted to equivalent circular radii, much like our own simulations (Section \ref{sec:fulldisk}). Like \cite{Makarov2009}, they fit spot and plage contrast to reproduce TSI, using a single contrast for all spots and a single \(\mu\)-dependent contrast for all plages. As with the other reconstruction studies, their typical jitter exceeds ours while their activity extremes are less pronounced. Additionally, they find no correlation between TSI variations and astrometric shift, in agreement with our analysis (Section \ref{sec:correlations}).}

\update{\cite{Shapiro2021} apply the SATIRE-S model to SOHO/MDI and SDO/MDI magnetograms. Unlike \cite{Makarov2010} and \cite{Lagrange2011}, they do not calibrate contrasts by fitting to TSI. Instead, spot and faculae contrasts are computed from the radiative transfer model ATLAS9, providing wavelength-dependant intensities for the following pixel classes: quiet Sun, umbra, penumbra, and faculae. Unlike \cite{Lagrange2011}, surface features retain the morphology of the surface features based on their magnetic field, rather than being approximated as circles. For the full 330--1050\,nm range, they report an astrometric jitter of \(1.12\mu \text{as pc}\) during high-activity periods and \(0.19\mu \text{as pc}\) during low-activity periods. In their blue (330--680\,nm) and red (680--1050\,nm) sub-bands, these become $1.30$ and $0.95\mu\text{as pc}$ (high activity) and $0.22$ and $0.16\mu\text{as pc}$ (low activity), respectively (Table~\ref{MainTable}). \cite{Shapiro2021} also provide scaling factors across Gaia passbands, finding a red/blue jitter ratio of 1.36 compared to our 1.56. This moderate difference likely reflects the breadth and range of the Gaia passbands relative to our narrowband filters.}

\subsubsection{Summary}
\label{ref:discussion_summary}
\update{To summarise, all three prior studies use reconstruction. \cite{Makarov2010} reconstructs from TSI-calibrated pixel classification, \cite{Lagrange2011} from catalogue positions with contrasts fitted to TSI, and \cite{Shapiro2021} from SATIRE-S modelling. In contrast, our approach directly measures the photometric intensity distribution of the solar disk without modelling surface features. The consistent pattern across comparisons is that our typical and quiet-period jitter is lower, while our high-activity jitter is comparable to or higher than reconstruction estimates, despite our data covering a weaker solar cycle. This wider dynamic range likely reflects a combination of methodological differences and wavelength coverage effects, as discussed in the preceding subsections. The low quiet-period values reinforce the importance of scheduling exoplanet searches during stellar activity minima where possible, as this would improve both detection probability and constraints on planetary parameters such as mass. Long duration missions will sample a range of stellar activity conditions, requiring pipelines to be robust against changing noise levels.}

\update{To the best of our knowledge, no prior study has measured astrometric jitter in a chromospheric activity line. Our Ca~{\sc ii} measurement (\(\sim 7 \mu \text{as pc}\) typical) provides the first empirical quantification of astrometric jitter in an active line, demonstrating the order-of-magnitude wavelength dependence and allows for constraints on the contribution of active lines to broadband jitter estimates.}


\subsection{Implications for detectability and constraints of exoplanets}


\update{The primary purpose of estimating astrometric jitter is to constrain exoplanet detection limits imposed by stellar activity. \citet{Meunier2022} provides a comprehensive assessment of this impact, computing detection limits for 55 nearby FGK stars via blind-test injection recovery on realistic activity simulations developed in \citet{Meunier2019} and \citet{Meunier2020}. These simulations extrapolate solar-like spot and plage distributions to spectral types F6-K4, and have been constrained by photometric and RV variability. For a G2 star (such as the Sun) viewed edge-on, their simulations yield a typical jitter of $1.06-2.66\mu\text{as pc}$ (Table \ref{MainTable}), higher than our directly measured solar value. The original THEIA detection limits \citep{TheiaPaper} assumed the equatorial solar jitter from \citet{Lagrange2011}, with \citet{Meunier2022} updating the limits with their new simulations. Since our typical jitter values in both the red and blue bands are lower than both of these estimates, we predict the stellar activity contribution to the noise budget is overestimated in that analysis. If a similar offset holds for other spectral types, their detection limits may be somewhat conservative, particularly for the nearest stars where the contribution of stellar activity is most significant. However, the detection limits in \citet{Meunier2022} depend on the full frequency structure of the stellar signal, so the actual improvement cannot be quantified without rerunning the injection-recovery simulations of that work calibrated with empirically measured jitter inputs.}

\subsection{Information contained within the jitter}

\begin{figure}
\centering
\includegraphics[width=1\linewidth]{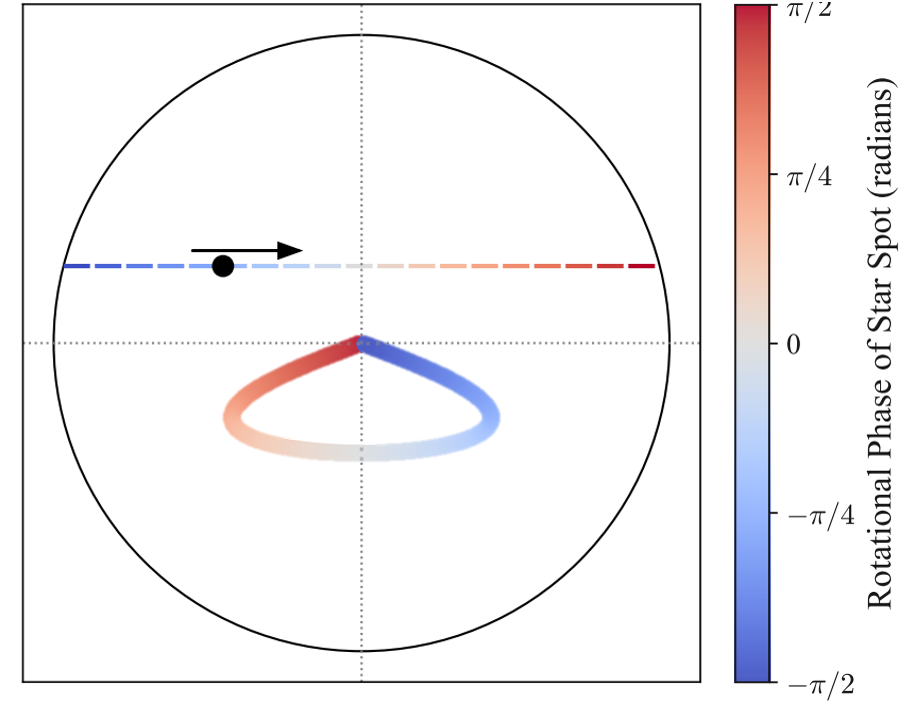}
\caption{A diagram of the shape of the photometric deflection caused by a single starspot rotating at a constant rate across a stellar disk, from \(-\pi/2\) (left limb / bluer) to \(+\pi/2\) (right limb / redder). The signal includes foreshortening and limb-darkening. The size of both the signal and starspot are not to scale.}
\label{fig_StarSpotSig}
\end{figure}

\begin{figure*}
\centering
\includegraphics[width=0.9\linewidth]{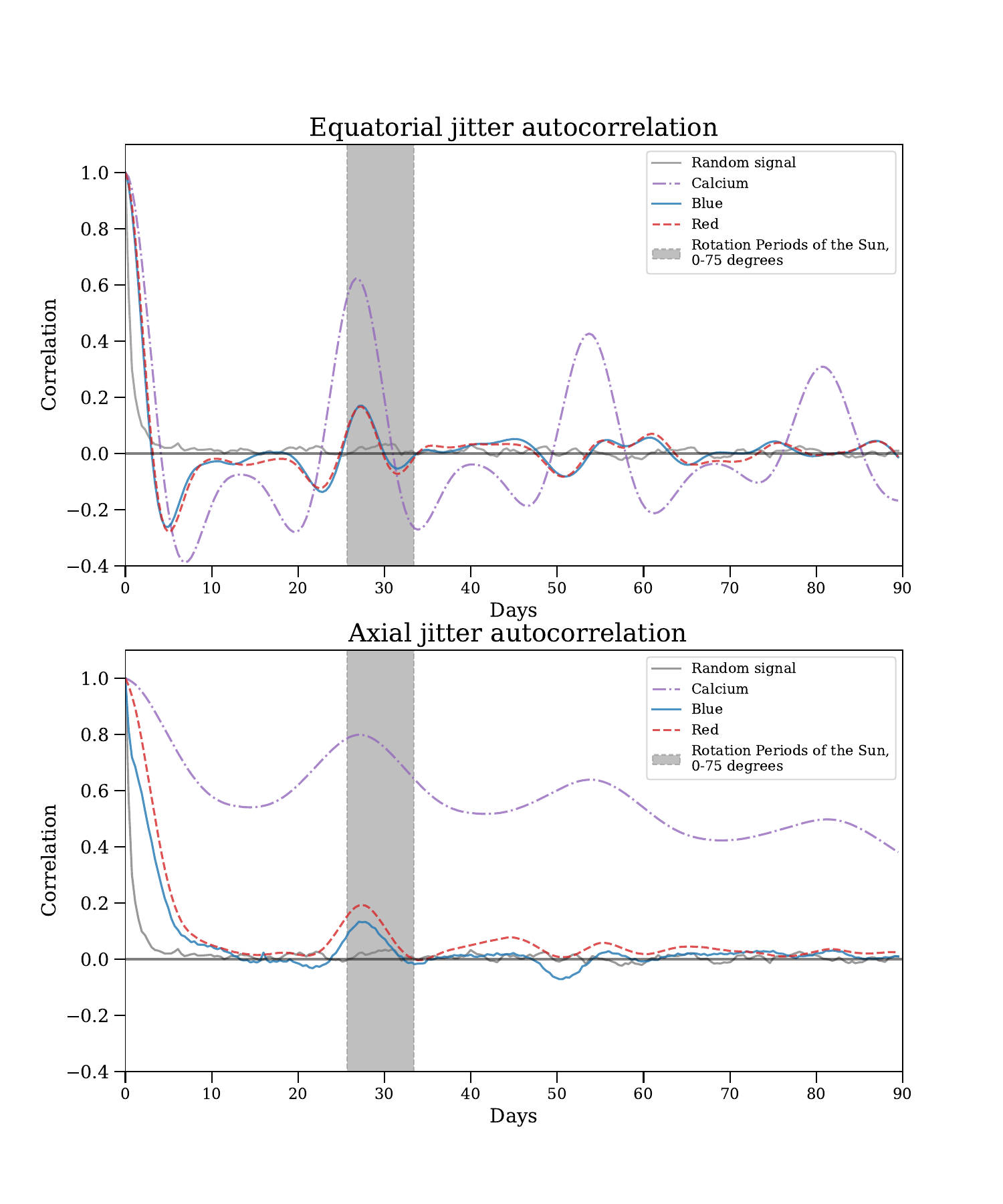}
\caption{The auto-correlation of the equatorial and axial (polar) astrometric signals, for the three PSPT wavelengths. The rotation period of the Sun is indicated by the grey shaded region. Given that the data was not uniformly sampled, we linearly interpolate the data in order for the autocorrelation to be performed. To ensure that this interpolation did not induce spurious structure in the above figure, we also perform a linear interpolation on random noise, as indicated by the grey line. Any structure that consistently deviates from this line can be taken as real structure.}
\label{fig:autocorr}
\end{figure*}

Astrometric jitter is not a random noise source, but a signal primarily generated from large scale magnetic surface features, notably starspots (Figs \ref{fig_PictureInPicture} and \ref{fig_SimPictureInPicture}). The signal is evident in the relatively strong coherence time seen in Figure \ref{fig:autocorr}, which shows the linear autocorrelation of the signal. However, the structure of this signal is curvilinear, as demonstrated in Figure \ref{fig_StarSpotSig}, so the autocorrelation does not give the full picture of the signal present.

The size, shape, \update{and} contrast of spots, as well as the number of spots present, all influence the magnitude and shape of the stellar astrometric signal, as demonstrated by \cite{Shapiro2021}, \cite{Sowmya2021}, and \cite{Lagrange2011}. It is therefore reasonable to assume that the astrometric signal thus contains information about the tomography of the stellar surface, which could be retrieved in a manner akin to light curve inversion (LCI). In fact, this method would contain more information than light curve inversion and could break some degeneracies that plage LCI (\cite{Luger2021}). 

The idea of using astrometric signals to infer information about the star itself is not new, and has been suggested by \cite{Lagrange2011} and \cite{Morris2018}. The full extent of retrievable stellar information via astrometry (tomographic or otherwise) will be explored in a forthcoming paper \update{\citep{Deagan2026}; see also \citet{Taaki2026} for a complementary approach.)}

\section{Conclusion}
\label{sec:conclusion}
The astrometric jitter of the Sun, a factor in detecting Earth-like exoplanets via astrometry, varies significantly with wavelength and solar activity cycle. Our analysis of high-resolution PSPT data spanning nearly a decade (2005--2015) provides several insights into this phenomenon:

\begin{enumerate}
    \item \textbf{Implications for exoplanet detection}: Most importantly, we find that even during high solar activity, the astrometric jitter of the Sun (1.294 $\mu$as pc at maximum) remains below the expected $\approx$3 $\mu$as astrometric signal that an Earth twin would produce around a Sun-like star at 1 pc. This suggests that for Earth-like exoplanet searches around Sun-like stars, within 2.5 pc  (750 pc for Jupiter-mass planets with a semi-major axis of 1 AU) instrument precision---not stellar activity---will be the limiting factor.

 \item \textbf{Wavelength dependence}: We found that the astrometric jitter varies substantially across different wavelengths. The Ca\,II K line (393.4 nm) shows jitter approximately an order of magnitude larger than in the red (607.2 nm) and blue (409.3 nm) filters. For future astrometric missions, we recommend: (1) avoiding chromospheric activity lines (e.g., Ca\,II, H$\alpha$, or photospheric iron lines) by employing filters that exclude these wavelengths, for instance through the use of notch filters. This approach would be akin to the successful line masking strategy of \cite{Dumusque2018} for reducing stellar RV jitter; (2) prioritising observations in red passbands where spot-to-photosphere contrast is minimised; and (3) using multiple filters at different wavelength ranges to constrain the impact of stellar activity on astrometric measurements, taking advantage of the changing contrast of spots relative to the photosphere across the spectrum.

    \item \textbf{Solar cycle variation}: The jitter magnitude changes by an order of magnitude between solar maximum and minimum. At $\lambda = 607.2$ nm, we measured typical jitter of \update{0.342} $\mu$as pc, ranging from \update{0.058} $\mu$as pc during solar minimum to \update{1.294} $\mu$as pc during maximum activity periods.
    
    \item \textbf{Comparison with previous work}: Our results  \update{show a wider range than} previous estimates, with lower jitter during quiet periods and higher \update{peak} jitter during active periods.  \update{The magnitude of these differences varies by comparison study and waveband; year-averaged values show closer agreement with the wavelength-resolved estimates of \citet{Shapiro2021} despite differences in Solar Cycle Strength, while 90-day extremes diverge more strongly. This pattern likely stems from a combination of wavelength coverage and methodological differences between direct photometric measurements and reconstruction-based approaches.}
    
\end{enumerate}

These findings support the viability of future high-precision astrometric missions like TOLIMAN and SHERA for detecting habitable-zone terrestrial planets. However, our results also highlight the importance of observations during stellar activity minima to maximise detection probability, as well as careful wavelength selection to minimise activity-induced noise.

Future work will explore how this astrometric jitter can be leveraged to extract information about stellar surfaces, potentially enabling astrometric tomography of distant stars and helping to distinguish between planetary signals and stellar activity in challenging detection scenarios.

\section*{Acknowledgements}

\update{This work was supported in part by the Breakthrough Prize Foundation as a part of the Breakthrough Watch Initiative.}

This research includes computations using the computational cluster Katana supported by Research Technology Services at UNSW Sydney \citep{katana}.

\update{This research was supported by the Commonwealth through an Australian Government Research Training Program Scholarship [DOI: https://doi.org/10.82133/C42F-K220]}

\section*{Data Availability}

The PSPT data used in this work is publicly available at \url{https://lasp.colorado.edu/pspt_access/}. Additional data is available upon reasonable request



\bibliographystyle{mnras}
\bibliography{example} 






\bsp	
\label{lastpage}
\end{document}